\documentclass[12pt,preprint]{aastex}
\def\cha{{\sl Chandra }}

\shorttitle{{\bf \cha} 
Observation of the  Shell of Classical Nova Persei (1901)}
\shortauthors{Balman \c{S}.}                                                

\begin{document}
\title{The {\bf \cha} Observation of the 
Shell of Nova Persei 1901 (GK Persei): Detection of localized 
Non-thermal X-ray Emission from a Miniature Supernova Remnant}
 \author{ \c{S}\"olen Balman}
\affil{Middle East Technical University, In\"on\"u Bulvar{\i}, Ankara, Turkey, 06531; solen@astroa.physics.metu.edu.tr}
\begin{abstract}

I present the data of 
the shell of classical Nova Persei (1901) obtained by the Advanced 
CCD Imaging Spectrometer S3 detector on-board \cha Observatory. 
The X-ray nebula is affected mostly by the complex interstellar
medium around the nova and has not developed a regular shell.
The X-ray nebula is lumpy and asymmetric with bulk of emission
coming from the southwestern quadrant. The brightest X-ray emission is detected
as an arc that covers from the west to the south of the central
source. Part of this feature, 
which is co-spatial with the brightest non-thermal radio
emission region, is found  to be a source of non-thermal (synchrotron) X-ray emission
with a power law photon index of 2.3$^{+1.5}_{-0.9}$ and $\alpha$=0.68$^{+0.03}_{-0.15}$
at about a flux of 
1.7$\times 10^{-13}$ erg cm$^{-2}$ s$^{-1}$. This confirms that
the shell is a cite of particle acceleration, mainly in the
reverse shock zone. There are strong indications for 
nonlinear diffusive shock acceleration occurring in the forward 
shock/transition zone
with an upper limit on the non-thermal X-ray flux of 1.0$\times 10^{-14}$ erg cm$^{-2}$ s$^{-1}$.
The total X-ray spectrum of the nebula  consists of two 
prominent components of emission (other than the 
resolved synchrotron X-ray emission).
The component dominant below 2 keV is most likely a 
non-equilibrium ionization thermal plasma of 
kT$_s$=0.1-0.3 keV with an X-ray flux of 1.6$\times 10^{-11}$
erg cm$^{-2}$ s$^{-1}$.
There is also a higher temperature,
kT$_s$=0.5-2.6 keV, embedded, N$_H$=(4.0-22.0)$\times 10^{22}$ cm$^{-2}$,
emission component prominent above 2 keV. The unabsorbed X-ray flux from this
component is 1.5$\times 10^{-10}$ erg cm$^{-2}$ s$^{-1}$.
The X-ray emitting plasma is of solar
composition except for enhancement in the  
elemental abundances (mean abundances over the remnant) 
of Ne/Ne$_{\odot}$ and N/N$_{\odot}$
in a range 13-21 and 1-5, respectively.  
A distinct emission line of neon, He-like Ne IX, is detected which
reveals a distribution of several emission knots/blobs
and shows a cone-like structure with wings extending toward 
NW and SE at expansion 
velocities about 2600 km s$^{-1}$ in the X-ray wavelengths. 
 
The emission measures yield an average electron density in a range 
0.6-11.2 cm$^{-3}$
for both of the components (filling factor=1). The electron 
density increases to higher values $\sim$ 300 cm$^{-3}$ if the filling factor 
is decreased substantially.  The mass in the X-ray 
emitting nebula is (2.1-38.5)$\times 10^{-4}$ M$_{\odot}$.
The X-ray luminosity of the forward shock $\sim$ 4.3$\times 10^{32}$\ 
erg s$^{-1}$ indicates that it is adiabatic. The shocked mass, 
the X-ray luminosity and comparisons with other wavelengths
suggest that the remnant has started cooling and most likely is in
a Sedov phase.

\end{abstract} 


\section{INTRODUCTION}

Nova Persei 1901 (GK Per) is one of the most extensively observed and
studied classical nova shells over the entire electromagnetic spectrum.
It is known to be a fast ONeMg nova with V$_{eject}\simeq$ 1200 km s$^{-1}$,
M$_{eject}\simeq$ 7 $\times 10^{-5}$ M$_{\odot}$ and
distance $\simeq$ 470 pc (Payne-Gaposchkin 1957; 
Mc Laughlin 1960, Pottasch 1959).
Nova Persei (1901) is the first recorded nova to show a light echo due to the
reflection of light off the nearby interstellar material (Kapteyn 1901).
Coudrec (1939) shows that the light echo region and the
reflection nebula detected later in 1917 (Barnard 1917; Oort 1951)
around GK Per are results of
large-scale circumstellar sheets of dust clouds.

The optical remnant is 103$\times$90 arcsec$^2$ (diameter$\sim$0.23 pc).
The images show high asymmetry and the remnant has evolved into
series of knots and filaments. The bulk of emission arises in the
southwestern quadrant indicating interaction between the nova ejecta
and the ambient gas (Slavin, O'Brien $\&$ Dunlop 1995, Seaquist et al. 1989).
Furthermore, the remnant is  detected with the Very Large Array (VLA) 
at 1.49 $\&$ 4.86 GHz as a non-thermal,
polarized radio source with a spectral index of -0.67 and a flux of 24 mJy at 1 GHz
(Reynolds  $\&$ Chevalier 1984).
In addition, the flux densities are 8.7, 20.6, 29, 33 and 38 mJy at frequencies
4.86, 1.49, 0.608, 0.408 and 0.327 GHz, respectively (Biermann, Strom, $\&$ 
Falcke 1995).
This shows the existence of
shocked circumstellar or interstellar material. 
The Infrared Astronomical Satellite (IRAS) observations
at 60$\mu$m and 100$\mu$m reveal a symmetric far-IR emission
region extending around the nova out to 6 pc (17$^{\prime}$ either side)
suggested to be
an ancient Planetary Nebulae associated with the binary (Bode et al. 1987; Bode, O'Brien, $\&$ Simpson 2004).
Recent optical wavelength observations of the vicinity of GK Per
show that some of this symmetric nebulosity is produced during the quiescent
mass-loss phase of the central binary because of the evolved nature of
the secondary (Tweedy 1995).
More recent IR observations also indicate that the IR emission within
17$^{\prime}$ of the source is of material originating from the secondary
(Dougherty et al. 1996). Scott, Rawlings, $\&$ Evans (1994) have discovered
symmetric blobs of 
CO emission bracketing the central object at about $\pm$ 200 arc seconds.
Overall, it is widely believed that the GK Per nebula behaves
like a young Supernova Remnant (SNR) in the pre-Sedov phase interacting with
its dense circumstellar medium.

Classical novae are a subset of cataclysmic variables which are interacting
binary systems hosting a main-sequence secondary (sometimes a slightly 
evolved star) and a collapsed primary component, a white dwarf (Warner 1995).
An outburst on the surface of the white dwarf as a result of a thermonuclear 
runaway in the accreted material causes
the ejection of 10$^{-3}$ to 10$^{-7}$ M$_{\odot}$ of material at velocities
up to several thousand kilometers per second (Shara 1989; Livio 1994; 
Starrfield 2001).
Though there has been no previous
detection of old classical nova remnants
in the X-ray wavelengths (evolving like Supernova remnants), classical nova
remnants have been detected in the hard X-rays (above 1 keV) as a
result of wind-wind interactions
{\it in the outburst stage}
(Balman et. al. 1998; Mukai $\&$ Ishida 2001; Orio
et al. 2001; Ness et al. 2003).
In general, the hard X-ray radiation above 1 keV
emitted from the hot shocked gas in nova remnants
should be a powerful diagnostic for the nature of mass-loss
mechanisms in classical nova outbursts together with the morphology,
the elemental abundances and the evolution of nova shells.
A long-sought insight into this issue has been gained with the discovery
of the remnant old shell of classical nova Persei (1901) in the X-ray
wavelengths. 

The X-ray shell around GK Per was first discovered in a 50 ksec exposure
with the ROSAT High Resolution Imager (HRI) (Balman $\&$ \"Ogelman 1999).
The X-ray  nebula extended to 46$^{\prime\prime}$ southwest,
60$^{\prime\prime}$ northwest, 52$^{\prime\prime}$ southeast, and
43$^{\prime\prime}$ northeast of the point source with an elliptical
shape and lumpy morphology.
The count rate of the shell was found as 0.01$\pm$0.001 c s$^{-1}$.
The estimated spectral parameters of the nova shell were:
N$_H$ $\sim$ 1.3$\pm$0.3$\times 10^{21}$ cm$^{-2}$, and
kT $\sim$ 0.16$\pm$0.03 keV ($\sim$2$\times 10^{6}$ K).
The implied unabsorbed X-ray flux was
F$_x$ $\sim$ 3.0$\times 10^{-12}$ erg cm$^{-2}$ s$^{-1}$
with L$_x$ $\sim$ 8.0$\times 10^{31}$ erg s$^{-1}$ (470 pc).
In general, what was detected by the ROSAT HRI were the condensations
in the shell. 
Since the ROSAT HRI did not have any adequate spectral resolution,
a \cha Advanced CCD Spectrometer (ACIS) observation was proposed. 

In this paper, I present the characteristics of the \cha spectrum, 
its components,
and  the detailed morphology of the shell. Finally, I discuss the
results in comparison with the observations in the other wavelengths
together with their implications on the novae theory and evolution
of the classical nova shells.

\section{THE OBSERVATION AND THE DATA}

GK Per and its vicinity is observed using the \cha (Weisskopf, O'dell, $\&$
van Speybroeck 1996) Advanced CCD Imaging
Spectrometer (ACIS; Garmire et al. 2000)
for a 95 ksec on 2000 February 10,
pointed 1 arc minute 
offset from the nominal point on S3 (the back-illuminated CCD)
with no gratings in use yielding a moderate non-dispersive energy resolution.
The data are obtained 
at the FAINT mode. \cha has two focal-plane 
cameras and two sets of transmission gratings that can be inserted in the 
optical path (HETG and LETG; High and Low
Energy Transmission Gratings). The ACIS is used either to take high resolution
images with moderate spectral resolution or is used as a read out device for 
the transmission gratings. ACIS is comprised of two CCD arrays, a 4-chip array,
ACIS-I (four front-illuminated CCDs); and a 6-chip array, ACIS-S (four 
front-illuminated and two back-illuminated CCDs). The spectral resolution
of ACIS arrays varies depending on the type of CCD in use, front or back 
illuminated CCDs, and the row numbers in the given CCD as a function of 
distance from the read out node or the aim point. 
The ACIS-S3 has a moderate spectral resolution E/$\Delta$E $\sim$ 10-30 (falls
to about 7 below 1 keV)
with an unprecedented angular resolution of 0$^{\prime\prime}$.49 per pixel
(half-power diameter).
The \cha X-ray Center (CXC) has
carried out the standard pipeline processing on the raw ACIS events, producing
an aspect-corrected, bias-subtracted, graded (limited to grade 02346) and
gain-calibrated event list (revision 2 data). In order to double check the
standard processing, "ACIS process events" thread is used to calibrate the
level 1 data using the necessary gain and "fef" files with the aid of CIAO 3.1
and the suitable CALDB (v.2.27-28). The 
data analysis is performed with  CIAO version 2.1, 2.2 in the preliminary 
conference proceedings 
Balman (2002a,b) and in this paper with versions 2.3 and 3.1
along with XSPEC version 11.2.0/11.3.0 and XIMAGE version 4.0 
for data preparation, spectral and spatial analysis.
Prior to the analysis, the data are also filtered from the
effects of the random flaring that occur during the observations
with the ACIS-S by excluding events that are 3$\sigma$ above the mean count 
rate in the light curve which reduced the exposure time to about 
81 ksec after cleaning. A preliminary analysis of the data could be found in
Balman (2002a,b). 

\cha Observatory provides good spatial resolution data with high sensitivity
thus, an 81 ksec observation provides an excellent opportunity to detect
faint X-ray sources in the given region of the sky. To utilize this, the
CELLDETECT algorithm (Harnden et al. 1984) 
is used to detect sources in the whole \cha field image. As a result,
16 new X-ray sources are detected over 6${\sigma}$ confidence level
above the background
(K\"upc\"u-Yolda\c{s} $\&$ Balman 2002).
Most of the sources are found as possible Extragalactic sources. Some
of them could
also be a Galactic Cataclysmic Variable, a Low-mass X-ray Binary or a cooling
Neutron Star. 

\section{THE CHANDRA IMAGE OF THE X-RAY NEBULA}

The bright X-ray nebula extends to
52$^{\prime\prime}$ South, 41$^{\prime\prime}$ North,
45$^{\prime\prime}$ West, and 38$^{\prime\prime}$ East
of the point source with an irregular shape, most counts coming from a
region centered at the SW. 
The count rate from the vicinity of the shell is 0.10$\pm$0.01 c s$^{-1}$.  
The count rate ratio of the
hemisphere centered around
NE to the one centered on the SW is 2:3. Figure 1 displays an exposure
corrected image of the 
nebula between 0.3 and 10 keV obtained with the \cha ACIS-S (S3).
The ACIS transfer streak is cleaned and the pileup PSF of the central source
is removed from the raw image.
The central source (the
binary system) is extracted by
modeling a two-dimensional PSF normalized to 1.65 c s$^{-1}$
(i.e., similar to the source count rate without pileup)
using ChaRT and MARX (version 4.08) (see also section 4.1 and 4.2). 
Figure 1 shows emission from the shell that is 2${\sigma}$ above the
background with a               
resolution of 0$^{\prime\prime}$.5 per pixel. A variable smoothing
is applied on the image using a Gaussian of 
${\sigma}$=0$^{\prime\prime}$.5-1$^{\prime\prime}$.
The shell is not well developed with an average  
surface brightness of (4.9$\pm$2.9)$\times 10^{-9}$
photons cm$^{-2}$ s$^{-1}$ arsec$^{-2}$.   
The X-ray nebula shows complex lumped morphology with structure
(eg., clump/blob/filament)  sizes
constrained by the smoothing performed on the image $\sim$ 1$^{\prime\prime}$-
2$^{\prime\prime}$.
Large scale features could be collections of smaller clumps/blobs or 
arcs and filaments which indicate radial elongations.
It has a central circular region brighter on the western
side and there is a rather faint conical emission region more extended compared 
with the central brighter part (see section 6).

\section{ANALYSIS ON THE CHANDRA SPECTRUM OF THE NOVA REMNANT}
 
Since the \cha observation was the first detection of the spectrum
of such a nebulosity around a classical nova, several different
models were tested on the data to understand the characteristics.
The search showed that no single model could be fitted to
the spectrum of the entire shell with a $\chi_\nu^2$ better than 8.
The spectrum of the entire nova shell showed a thermal plasma emission 
below 2 keV.
The harder X-ray tail of the spectrum above 2 keV was also 
best fitted with
a thermal plasma emission model.  Equilibrium or non-equilibrium ionization models
yielded similar fits. A power law model was ruled out
for the spectrum of the entire remnant because of the
non-physical spectral indices derived from the results in contrast with
the non-thermal (synchrotron emission) nature of the radio data (Balman 2002a).  
The spectrum was remarkably flat above 2 keV. 
Figure 2 shows the best fitted two-component emission model of VMEKAL
(Mewe, Gronenschild $\&$ van den Oord 1985; Liedahl et al. 1995) + PSHOCK
(Borkowski et al. 1996). The shell photons were extracted using an
annulus with an inner
radius of 12$^{\prime\prime}$.5 and an outer radius of 
67$^{\prime\prime}$.5. The background was also derived from an annular region farther 
out around the nova normalized to the source extraction region. The spectrum was 
calculated using a minimum of 50 counts per bin and the channels below 0.3 keV and 
above 9 keV  were excluded due to low statistics. 

The spectral parameters that were derived from the fits to the second component
resulted in a very high shock temperature in excess of 50 keV and shock speed 
larger than 5000 km s$^{-1}$ which were inconsistent with all the  measurements in the
other wavelengths. This problem was largely due to the excess photon contribution from the wings of the
PSF of the piled-up central source. Such a contaminating component resulted in 
significantly higher temperature and flux for the second component above 2 keV, and shifts in the central line 
energies (toward lower energies) 
of the lower temperature component below 2 keV. This excess emission compared with the observed PSF of the
central source also needed to be properly cleaned in order to study the image of the shell.  
Sections (4.2) and (4.3) describe the problem and the 
procedure/method used to clean and remove this effect of the PSF wings from the shell data 
(i.e, images and spectra).
  
\subsection{The Contamination of the Shell Image by the Central Source}

Pileup results in the CCDs when two or more photons are detected as a single event.
The consequences are spectral hardening since apparent energy is approximately
the sum of two (or more) energies and underestimation of
the true counting rate of the point sources. 
It is expected that pileup problem also distorts the PSF of the
central source by creating holes at the centers (pulse saturation) and the PSF wings
become more prominent as the count rates increase.
As a result, appreciable central source counts can be 
detected in a region within 2$^{\prime}$ of the pointed source. This 
contaminates any extended emission in the close vicinity of the central 
source. The source photons of GK Per 
has a pileup of about 89$\%$ where one expects
the emission from the shell to be affected by the pileup PSF and the prominent
PSF wings. In order to exclude the spectrum of the central source, GK Per, the nova shell photons are extracted
from an annular region of 12$^{\prime\prime}$.5 to 67$^{\prime\prime}$.5 in radius (as presented in sec 4.0). 
This should largely avoid
the contamination from the piled-up central source, but it will be inefficient to exclude the effects of the
prominent PSF wings beyond 12$^{\prime\prime}$.5 out to 67$^{\prime\prime}$.5 in radius.
To segregate the counts of the nova shell 
from the central source counts in the wings of the PSF (at the vicinity of the nova shell), 
an archival search within the ACIS-S pointings
(no grating or HETG observations) has been performed. The prime motivation of the search has been to
measure the count rate in the PSF wings in an annular region of 12$^{\prime\prime}$.5 out to 
67$^{\prime\prime}$.5 radius in comparison with the count rate in a circular region of
12$^{\prime\prime}$.5 . The particular extraction radii is
assumed only to compare the GK Per observation with other archival observations.
The two different regions assumed are designated as: (1) the {\it source} defined as 
the circular photon extraction region of 12$^{\prime\prime}$.5 radius (25 pixels), and (2)
the {\it shell} defined as the annular photon extraction region of 
12$^{\prime\prime}$.5 to 67$^{\prime\prime}$.5 in radius. 
The count rates within the {\it source} and 
{\it shell} regions are compared for several observations so that an acceptable estimate of the
count rate in the wings of the PSF can be made for the central source of GK Per
in the vicinity of the nova shell given the central source count rate of GK Per (pileup rate). 
Over 300 archival data of {\it pointed sources} on the ACIS-S array have been analyzed 
imposing several criteria :(1) energy spectrum of the source peaks about 
1-2 keV and have a hard X-ray tail, (2) pointings are
within a region of 2$^{\prime}$ of the nominal point on ACIS S-3, (3) all
observations have the same frame time of 3.2 sec and obtained in the Faint mode, (4) no special
spatial filtering has been performed during the observation for pileup 
mitigation, (5) HETG observations with bright first order data have also been
excluded, (6) observations with pileup count rates brighter than a few c/s 
(i.e. 3 c/s) are also excluded since the scaling relation is altered, (7) 
observations that show strong background flares and anomalous high background rates have been disregarded. 
As a result, about 15 suitable pointings have been recovered
that fits all the criteria.  
Figure 3a shows the count rate in a
region of 12$^{\prime\prime}$.5 radius (25 pixels), the {\it source},
versus the count rate in an
annular region of 12$^{\prime\prime}$.5 to 67$^{\prime\prime}$.5 in radius,
the {\it shell}, both centered
at the point source.
Finally, a scaling relation between the count rates in two different
regions described as the {\it source} and {\it shell} is derived.
The curve in Figure 3a is best fitted
by the quadratic function, ax$^2$+bx+c, where a=-0.12$\pm$0.09,
b=1.11$\pm$0.14, and c=-0.18$\pm$0.03.
This has been used to approximate the count 
rate in the PSF wings of GK Per within the region 
covered by the classical nova shell. For a {\it source} count rate of 0.2 c s$^{-1}$ 
(pileup rate), the
estimated count rate in the PSF wings ({\it shell} region) within the classical nova shell is 0.037 c s$^{-1}$, which
yields an actual nova shell count rate around 0.063 c s$^{-1}$ (instead of
0.1 c s$^{-1}$). 
This is also checked using an archival ACIS-S HETG data of GK Per obtained during a
dwarf nova outburst. The zeroth order pileup source count rate is used to 
estimate the count rate in the described shell with the aid of the derived
scaling relation (same extraction radii is assumed). Considering that a lower frame time, and spatial clipping are
used to mitigate the pileup to some extend and that the source is in outburst, the scaling function yields similar
results. I would like to note that the contamination problem described here
could be important for \cha ACIS observations of other astrophysical objects 
like Galaxies/Clusters and centrally filled composite SNRs. 

In order to remove the central source from the image of the nova shell of GK Per together with the prominent PSF wings,
a PSF is created using ChaRT. 
Several PSFs are merged to achieve the correct count rate within the
nova shell region (i.e., 0.037 c s$^{-1}$ in the PSF wings). 
The final central source count rate of the combined PSF in the {\it source} region is close to 
the  count rate (within 8$\%$) of the 
central source, GK Per, when there is no pileup (1.7-1.8 c s$^{-1}$). The un-piled central source rate is
calculated using an archival ASCA data of GK Per (i.e., spectrum) and PIMMS.  
Next, the combined PSF is extracted from the total image to achieve Figure 1.     
 
\subsection{The Cleaned Shell Spectra}

The central source, GK Per, has 89$\%$ pileup and has about 0.037 c s$^{-1}$
in the region of the X-ray shell due to its prominent PSF wings. Thus, this will also affect
the "entire nova shell spectrum" where the spectrum of the central source will 
be appearing as a component of the nova shell spectrum of GK Per as a result of the counts
in the PSF wings.
I would like to note that though GK Per has a dusty
environment, previous X-ray observations with several different satellites have
not recovered any dust scattering halo around the source which could have affected
the emission below 1 keV.  
The result of the analysis presented in section (4.1), for
the sources which suffer from the pileup effect on ACIS-S3, indicates that the spectra
derived from the {\it shell} regions (as denoted in section [4.1]), basically from the 
PSF wings, may be better approximated with the spectra 
derived from the {\it source} regions, but scaled down. This is only because the spectral shapes resemble
each other, and not that the spectra from the PSF wings have pileup photons (as in the {\it source} regions). 
Theoretically, pileup in the PSF wings
should either be non-existent or very small,
however the spectra from the PSF wings may show spectral hardening. 
Figure 3b shows a collection of  
spectra of the sources used for calibration where the above assumptions can be clearly investigated.
The upper curves represent the {\it source}-region spectra and the lower curves  represent the
{\it shell}-region spectra (spectra obtained from the PSF wings). The sources have 
very similar pileup source count rates to GK Per in the stated {\it source} extraction region 
0.19-0.21 c s$^{-1}$ (more or less similar central source spectra with a hard X-ray tail - three objects 
are other CVs).
In order to remove the spectrum of the PSF wings of the central source from the nova shell spectrum, 
first a  spectrum (of GK Per) 
is extracted from a circular region of radius 12$^{\prime\prime}$.5 ({\it source}-region). This  spectrum
is fitted with a composite model of photoelectric absorption, bremsstrahlung and power law models. 
The best fit parameters are used to
recalculate/rescale this spectrum using both MARX and WEBSPEC yielding a count rate of 0.037 c s$^{-1}$.
The parameters used to model this {\it source}-region spectrum are an N$_H$ of 1.16$\times$10$^{22}$, a
covering fraction of 0.95, a kT of 13.3 keV, a bremsstrahlung normalization of 
5.54$\times$10$^{-4}$,
a photon index of 2.5 and a power law normalization of 7.93$\times$10$^{-7}$.
Figure 4 shows the central source spectrum of GK Per derived from the {\it source}-region (upper curve), 
the modeled spectrum to be subtracted
(middle curve), and the un-piled central source spectrum of GK Per (lower curve). The un-piled source spectrum
is extracted from the out-of-time events of the ACIS transfer streak.  I want to point out that
all three data sets in Figure 4 are consistent with one another and  the un-piled source spectrum
can be approximated with the modeled spectrum to an acceptable extend.
Finally, the modeled spectrum (middle curve) is subtracted PI channel by PI channel from the nova shell spectrum of GK Per.
Figure 5a shows
the resulting total nova shell spectrum after removing the spectrum of the 
PSF wings of the central source, GK Per. 
The total number of nova shell photons between 0.3-1.6 keV is 4675 and between 1.6-8.0 keV is 
410 after removal of the background and the photons from the wings of the central source PSF. 
The shell photons are extracted using an annulus with an inner radius of 12$^{\prime\prime}$.5 
and outer radius of 67$^{\prime\prime}$.5  and regrouped using a signal-to-noise ratio of 13
per energy bin. 
The cleaned nova shell spectrum can not be fitted with a single emission component
with reduced  $\chi_\nu^2$ of 4.0 (see Figure 5a).
{\it There is excess emission above 1.6 keV that causes up to  4$\sigma$ deviations
in the residuals when the data are fitted with a single spectral model}. 
In the rest of the paper, only analyses and results derived from the cleaned spectra/image will be
presented, since there are significant differences between the cleaned and uncleaned data.

\subsection{The First Component (below 2 keV)}

The X-ray emission below 2 keV is found to be consistent with only
the thermal plasma emission models. Figure 5b shows the data and the 
best fitted two-component emission model of VPSHOCK+NEI (Borkowski et al. 1996;
Hamilton et al. 1983). The shell photons are extracted using an annulus with an inner radius of
12$^{\prime\prime}$.5 and outer radius of 67$^{\prime\prime}$.5. The background
is also derived from an annular region farther out around the nova normalized to the 
source extraction region. The spectrum is calculated using a minimum of 50 counts
per energy bin (or a Signal-to-noise ratio of 13 per energy bin, which made no difference
in derived parameters). The PI channels below 0.3 keV and above 8 keV are excluded due to low
statistical quality.    
The spectral parameters of the nova shell are 
an N$_H$ of (0.1-3.7)$\times 10^{21}$ cm$^{-2}$,
kT of 0.1-0.3 keV ($\sim$(3-4)$\times 10^{6}$ K) and an
emission measure (EM) in a range (0.2-54.5)$\times 10^{54}$ cm$^{-3}$. 
The noted error ranges correspond to  2$\sigma$ confidence level.  
The spectral
parameters derived for this component using three different composite
spectra are, also, displayed on Table 1. 
The unabsorbed soft X-ray flux is
F$_x$ $\sim$ (0.03-11.8)$\times 10^{-11}$ erg cm$^{-2}$ s$^{-1}$
yielding an X-ray luminosity of
L$_x$ $\sim$ 4.3$^{+26.1}_{-4.2}$$\times 10^{32}$ erg s$^{-1}$.
Any two-temperature (T$_i$$\ne$T$_e$) non-equilibrium ionization plasma
model yields similar T$_i$ and T$_e$ within 2$\sigma$ error ranges when fitted to the data below 2 keV.

The detected emission measure EM=$<n_e>^2$ V$_{eff}$ (after cleaning the shell spectrum) 
yields an average electron density n$_e$ in a range 0.6-11.2
cm$^{-3}$ using a volume of 4.3$\times 10^{53}$ cm$^{3}$ (consistent
with the X-ray photon extraction region) and  a filling factor of 1
(i.e., V$_{eff}$ can be theoretically expressed as V$_{eff}$=1.33$\pi$v$^3$t$^3$f; v 
is the expansion velocity [see sec (1)], t is the elapsed time 
since the eruption, and f is the volume filling factor). 
If the filling factor is as low as 
1$\times 10^{-5}$,  
then the electron density can be as high as 84-712 cm$^{-3}$ ($\sim$ 300 cm$^{-3}$). 
The time scale for equipartition between electrons and ions is
\begin{equation}
t_{e-i} \simeq 2.5\times10^6  \left({T_e \over 10^9K} \right)^{1.5} \left({n_e \over 10^8cm^{-3}} \right)^{-1}
\end{equation}
(Spitzer 1978; Fransson, Lundqvist $\&$ Chevalier 1996, hereafter FLC96).
The derived range of the X-ray temperature (T$_e$$\sim$T$_{shock}$) 
and the age of the remnant 
implies that the emitting plasma is in (for n$_e$$\ge$60 cm$^{-3}$) 
or close to equilibrium. The ionization timescale ($\tau$=n$_0$t) of $\tau$$<$1.7$\times 10^{11}$ 
detected using the PSHOCK (non-equilibrium ionization plasma) model indicates 
it is close to equilibrium with ambient density n$_0$$\le$ 60 cm$^{-3}$ (t=99 years).
This is supported by the HI measurements in the vicinity of
GK Per where the column density infers an ambient density of 20-30 cm$^{-3}$
(Seaquist et. al. 1989).  The filling factor in the region where the first component
arises is f$>$0.0022 calculated using n$_0$ (thus, n$_e$=4n$_0$) and the range of EM derived from the
spectral fits. 

The third composite model on Table 1 (i.e., \S{3})
is constructed to derive the detected line energies, line sigmas and fluxes together
with the continuum temperature.                     
The X-ray emitting plasma
has mostly solar composition except for neon and nitrogen. The
Ne/Ne$_{\odot}$ is in a range 13-21 and  N/N$_{\odot}$ is in a range 1-5 (ratio of
relative number fraction to H).
The neon over-abundance is detected in a prominent emission line
at around  E$_0$=0.900-0.916 keV (2$\sigma$ range) corresponding to
the position of the He-like Ne IX
emission line triplet with a flux F=(1.7-11.0)$\times 10^{-14}$
erg cm$^{-2}$ s$^{-1}$ (three separate lines are not resolved). 
The range of the central line energy support that the (He-like) Ne IX forbidden line 
emission may be more pronounced and that the first X-ray component is an under-ionized
plasma, a non-equilibrium ionization plasma, cooling toward  collisional equilibrium 
(the nonexistence of the H-like neon line will strengthen this). At around 3-4$\times 10^{6}$ K,
the neon is expected to start recombining.
Nitrogen over-abundance is highly likely resulting from the He-like N VI
emission line at
E$_0$=0.412-0.433 keV (2$\sigma$ range) with a flux
F=(0.35-3.16)$\times 10^{-13}$ erg cm$^{-2}$
s$^{-1}$.  The spectral resolution of ACIS degrades at such
energies and the line profile is smeared. The exclusion
of the line causes deviation above three sigma level at around the line 
energy (Balman 2001). The shift in the central line energy toward lower energies is also observed for the N VI
indicating a more prominent forbidden line emission.

The continuum temperature derived
from the thermal bremsstrahlung fit to the data below 2 keV is consistent
with the fits performed using the other models within errors. 
The two different Gaussians used to fit the data,
yield the Ne and N lines as mentioned above.
The third Gaussian has a central energy  E$_0$=0.549-0.562 keV (2$\sigma$ range)
which corresponds to the O VII emission line. This indicates that
the low temperature plasma is also rich in oxygen.  
A fit to the data with increased oxygen abundance yields an abundance
1-3 times the solar abundance of oxygen (ratio of relative number fraction to H). 
However, I need to caution that
the O VII line  overlaps with the maximum observed energy of emission of the low 
temperature component. 
An observation with the
LETG/HETG (i.e., grating observations to resolve separate lines) 
on board \cha to resolve the emission lines 
would be unfeasible for the shell because it has a very low
surface brightness and a more recent HETGS observation of the central source
shows no sign of the shell. 

\subsection{The Second Component}

The second component is found to be consistent either with a
thermal plasma emission or a bremsstrahlung emission in origin. 
The spectral parameters obtained from the fits with the 
PSHOCK model are an  
N$_H$ of (4.0-22.0)$\times 10^{22}$ cm$^{-2}$, a shock temperature
of kT=0.5-2.6 keV (best fit $\sim$ 0.9 keV) and an  emission measure of
(1.5-160.2)$\times 10^{54}$ cm$^{-3}$.
Table 2 displays the spectral parameters
for the entire shell derived from the three different composite model fits
to the X-ray data above 2 keV. The ranges correspond to 2$\sigma$
confidence level and the fits are carried out simultaneously with the first
component. Figure 5b shows a fitted composite spectrum.
A very intriguing outcome from the spectral fits is the detection of a cold shell
between the two components that shows high absorption from the neutral material within this zone.
To clarify this, a fit with a single N$_H$ parameter is shown in Figure 5c (see also  Table 2) where 
significant excess above 2 keV is observed with $\chi^2_{\nu}$ values of about 2.4 .
Moreover, the derived plasma temperature (using thermal plasma models) 
from fits with a single absorption (N$_H$) parameter is very large compared with expectations.
Thus, single absorption for both components of emission in the shell spectrum is non-physical. 
Fits with a non-thermal power law or SRCUT model also show non-physical results with  positive 
spectral/photon indices (see Table 2) for a single absorption parameter for the entire shell spectrum.
The detailed discussion on this issue can be find in section [7.1].
The ionization timescale (i.e., ${\tau}$=n$_0$t) is found as
 ${\tau}$=0.9$^{+41.0}_{-0.8}$$\times 10^{10}$ s cm$^{-3}$. 
The plasma shows no resolved emission lines.
On the other hand, the high neutral hydrogen column density
absorbs the X-ray emission below 1.5 keV where most of the
emission lines (e.g., Ne and N) would be found.
The unabsorbed X-ray flux of the high temperature component
is F$_x$ (0.5-25.0)$\times 10^{-11}$ erg
cm$^{-2}$ s$^{-1}$ consistent with an X-ray luminosity of
L$_x$=(6.6-0.2)$\times 10^{33}$ erg s$^{-1}$ (at 470 pc).
The reduced $\chi^2$ values do not differentiate between
equilibrium and non-equilibrium ionization plasma emission models.
The emission measure implies a minimum 
average electron density of 2.0 cm$^{-3}$ (maximum 19.3 cm$^{-3}$) 
over the volume stated in section (4.3) with a filling factor f=1.
If the filling factor is decreased to about 1$\times 10^{-4}$, the electron
density increases to about 200 cm$^{-3}$. 
The ionization timescale translates to ambient densities of n$_0$ $<$ 130.6 cm$^{-3}$ for the
second component with 4.1 cm$^{-3}$ calculated from the best fit results of the 
ionization timescale. 
Using the EM parameter values and scaling the electron density
from the maximum limit on the ambient density (4n$_0$=n$_e$), 
the filling factor is f $>$ 0.0002 . 

Non-thermal power law and SRCUT (Reynolds 1998; Reynolds $\&$ Keohane 1999) 
models have also been applied to the shell
spectrum above 2 keV. The resulting spectral parameters are 
an N$_H$ of (6.0-16.0)$\times 10^{22}$ cm$^{-2}$ (using two different  N$_H$ parameter), a photon index 
$\Gamma$=5.9, a spectral index $\alpha$=0.32 and an unabsorbed X-ray flux 
F$_x$ $\ge$ 8.0$\times 10^{-13}$ ergs s$^{-1}$ cm$^{-2}$. Table 2 shows
the detailed results of the fits using these models.
The photon index implies 
steeper power law spectra not in accordance with the findings on  
SNRs that indicate existence of particle acceleration in the X-ray and radio wavelengths
(see section [5.0]). Such values are in a range $\Gamma$=2-4 within errors 
(Rho et. al. 2002, Berezkho et al. 2003). The spectral index is also inconsistent with the
findings on the shell of GK Per in the radio wavelengths (see the Introduction).

\section{THE DETECTION OF LOCALIZED NON-THERMAL X-RAY EMISSION FROM
THE NOVA REMNANT AND SPECTRAL VARIATIONS}  
 
Once the whole spectrum of the shell is determined, spectral variations on
smaller scales are examined. The remnant is divided into four azimuthal
quadrants around the central source from position angle PA=0$^{\circ}$
to PA=360$^{\circ}$ centered on NW, NE, SE, and SW. Figure 6a shows a graphical
representation of the extraction regions.
The resulting spectra are simultaneously fit with a model similar to
the spectrum of the entire shell. The spectral
parameters show slight variations in different quadrants, but the variations
are within 2$\sigma$ confidence level errors in most cases. 
The four spectra have count rates in a range
0.019-0.006 c s$^{-1}$ (the spectrum of the PSF wings of the
central source is removed with a similar method as described in sections 4.1 and 4.2). 
There are variations in each spectra at different
energies below 1 keV (see Figure 6b). These could be related to 
variations in the line intensities (variation in elemental abundances could also be possible).
NE quadrant has the least line intensity between 0.3-1 keV 
(see the 4$\sigma$ variation between NE and SW
at around the energies corresponding to the O VII emission line in
Figure 6b). 
The asymmetric morphology of the X-ray nebula is the 
result of the distribution of material in the
circumbinary medium of the nova and geometrical effects.

The search for spectral variations over the entire remnant  shows that
the spectrum of the nebula with two thermal components is generally
consistent over the whole remnant. On the other hand, the 
radio synchrotron emission from the nova shell strongly suggests that 
non-thermal emission associated with the accelerated electrons
should be recovered. Such a component that is not prominent enough
to be detected from the entire shell, then could be localized within the
shell. Most of the efforts to derive spectral variation across the remnant suffered from
the low surface brightness of the shell except for the highest intensity zone
at the edge of the remnant
which is an arc of emission covering from NW to S of the X-ray nebula.
This is also co-spatial with  the brightest
radio synchrotron emission region at 1.4 GHz (see section [6.1]).    
The spectrum of the brightest X-ray emission region, {\it Peak Spectrum}, 
is extracted from the edge of the shell
using a sector that subtends  about 70$^{\circ}$ to the  central source
with a width of nine arc seconds (the region denoted as (2) in Figure 6a). Next,
the Peak Spectrum is cleaned using a similar procedure described in
section (4.1 $\&$ 4.2). In addition, 
the rescaled source spectrum is  corrected further 
using the total number of central source photons within the extraction annulus of the entire nova 
shell spectrum ($\sim$ 3000 counts)
and a radial distribution of r$^{-2}$ (i.e., N(r)$\sim$Ar$^{-2}$\ see \cha calibration pages
for further details).   
The simulated central source spectrum (as described in section 4.2)  
is subtracted PI channel by PI channel from the Peak Spectrum. The cleaned Peak Spectrum
is regrouped with a minimum of 20 counts per energy bin. 
The uncleaned Peak Spectrum has a count rate of 0.0099$\pm$0.00072 which reduces
to 0.00925$\pm$0.00057 after the cleaning process is completed. It is apparent that
the cleaning process is not crucial and necessary for the bright X-ray emission region
since this region is far away from the central source and small enough so that the
contribution from the PSF wings of the central source is marginal (below the local background).
The total number of photons between 0.3-1.6 keV is 670 and between 1.6-8.0 keV is 105 after the
removal of the background and the photons resulting from the wings of the central source PSF.
Because of the low statistical quality PI channels below 0.3 keV and above 8.0 keV have been excluded.

An investigation of the cleaned Peak
Spectrum, performing fits to the data reveals that it has two components. 
Table 3 displays all the results from
fits  with several composite thermal and non-thermal emission models.
Figure 7a shows the Peak Spectrum fitted with a single model of emission where
the reduced $\chi$$^2_{\nu}$ value is 3.4 and the excess in the band 1.6-8.0 keV
is significant. Thus, existence of two components of
emission is justified.
The first component resembles to that
of the first component of the entire shell with similar spectral parameters.
The spectrum of the second component from the X-ray Peak region, this time, 
yields consistent fits
with the SRCUT or power law emission models as opposed to a thermal plasma model which yields
non-physical results. All the fits thermal plasma models result in temperatures 
in excess of 7 keV (2$\sigma$ lowest limit and best fit result $\sim$ 20 keV) which is
inconsistent with the expectations from this remnant. The fits also reveal absorption
difference between the two components in question and the fits involving a single absorption
parameter for the two components yield significant excess in the 1.6-8.0 keV range with reduced $\chi$$^2_{\nu}$
value of 3.0 .
Figure 7b and Figure 7c   
display examples of the fitted Peak Spectrum that has a two-component model with
two different  N$_H$ parameter, and
a two-component model with a single  N$_H$ parameter for both components, respectively.
The spectral parameters for the second component derived using a simple power law model
are an N$_H$ of 4.9$^{+5.1}_{-3.7}$$\times 
10^{22}$ cm$^{-2}$, a photon index of $\Gamma$=2.3$^{+1.5}_{-0.9}$
and an unabsorbed  X-ray flux of 1.7$^{+73.0}_{-1.4}$$\times 10^{-13}$ erg s$^{-1}$ cm$^{-2}$. 
The errors correspond to 2$\sigma$ confidence level. 
The X-ray flux  of this non-thermal component
translates to an unabsorbed X-ray luminosity of about
4.6$\times 10^{30}$ erg s$^{-1}$. More appropriate models for
synchrotron emission (from SNRs) involve exponentially cut-off power law
distribution of electrons in a magnetic field (Reynolds 1998; Reynolds $\&$ Keohane 1999).  
This is largely because the synchrotron losses or escaping particles are expected to
decrease the flux density at the X-ray energies falling below the extrapolated radio spectrum.
The cleaned spectrum from the X-ray Peak region (region [2] in Figure 6a) is fitted with the 
SRCUT model for a more proper treatment.
The resulting parameters are : N$_H$=5.0$^{+5.5}_{-1.7}$$\times10^{22}$ cm$^{-2}$; 
$\alpha$=0.68$_{-0.15}^{+0.03}$; and $\nu$$_{break}$=1.2$^{+2.3}_{-0.3}$$\times10^{18}$ Hz.
I would like to make a note that since the photon extraction region for the Peak Spectrum is very small
and farther away from the central source, the cleaning process have slightly
changed (i.e., corrected) the best fit parameters leaving the error ranges similar/unaltered 
(for both SRCUT and the power law models).
The normalization parameter, which is the radio energy flux at 1 GHz (in Jy), is fixed at the calculated value
of 0.024 Jy (Reynolds $\&$ Chevalier 1984; Biermann et al. 1995). 
The energy flux and the unabsorbed X-ray luminosity
is the same as the values calculated for the simple power law model. The fit successfully recovers
the correct spectral index ,$\alpha$, for the radio spectrum ($\alpha$=0.67; see Introduction).    
This is also consistent with the range of X-ray spectral 
photon index. {\it Thus, emission from the same electron population 
has been detected as the extension
from the radio wavelengths.}   

It is expected  to detect gamma-ray emission as well as X-rays from the shell-like SNRs
primarily due to bremsstrahlung of accelerated relativistic
electrons off pre-shock material together with
contribution from an Inverse Compton emission
(off of Cosmic Microwave Background), the
pion decay interactions, and the secondary  p-p interactions (Berezhko $\&$ V\"olk 1997;
Grassier et al. 1998; Stunner et al. 1997; Baring et al.
1999, Berezkho et al. 2003 and references therein). 
The critical frequency (also the break frequency in the above paragraph) for
synchrotron emission from an electron of energy E is
\begin{equation}
{\nu_{break}} = 3.2\left({B sin \theta \over 2\times10^{-4} G} \right) \left({E \over GeV}\right)^2 GHz
\end{equation}
(Chevalier 1999). Using the cut-off (break) frequency derived from the SRCUT model fits,
I estimate E$_{max}$ $\simeq$ 15-30 TeV for the
maximum energy of accelerated electrons radiating at (0.4-1)$\times 10^{18}$ Hz for the remnant of
GK per (B=75-47$\mu$G; Seaquist et al. 1989).

A spectrum is extracted from the outer ridge (outer ridge: region denoted as (3) in Figure 6a) using a PIE extraction
region with a width of nine arc seconds (the same size annulus-sector 
as the photon extraction region of the Peak Spectrum). 
A fit with the SRCUT model of emission
yields an N$_H$=2.2$^{+1.8}_{-0.7}$$\times10^{22}$ cm$^{-2}$,
$\alpha$=0.76$_{-0.02}^{+0.03}$ and $\nu$$_{break}$$\ge$1.8$\times10^{18}$ Hz ($\chi$$_{\nu}^2$=1.17 (37)). 
The unabsorbed
X-ray flux from the outer ridge is 4.1$_{-2.2}^{+2.0}$$\times10^{-14}$ erg s$^{-1}$ cm$^{-2}$. 
 The spectrum of the inner ridge (inner ridge: region denoted as (1) in Figure 6a) is extracted,
using a similar photon extraction region as in the derivation of the 
Peak Spectrum and the outer ridge (with a width of nine arc seconds). 
A fit with the SRCUT model of emission
yields an N$_H$=2.1$^{+3.5}_{-1.3}$$\times10^{22}$ cm$^{-2}$,
$\alpha$=0.58$_{-0.07}^{+0.07}$ and $\nu$$_{break}$=3.1$_{-2.1}^{+4.0}$$\times10^{16}$ Hz 
($\chi$$_{\nu}^2$=0.7 (23)). 
The unabsorbed 
X-ray flux from the inner ridge is 1.1$_{-0.07}^{+5.2}$$\times10^{-13}$ erg s$^{-1}$ cm$^{-2}$.
All the spectra are cleaned (for the wings of the PSF of the central source) 
using a similar procedure to the Peak Spectrum where the amount of cleaned PSF photons is
largest for region (1) and lowest for region (3). 
The fits with the SRCUT model for the regions 1, 2 and 3 in Figure 6a 
yield spectral indices consistent with the radio observations of the remnant which are  found to be
in a range 0.55-0.95 (between 1.4-4.9 GHz; see Seaquist et al. 1989). 
The radio spectral index is variable over the remnant where it is steepest at the
outer ridge and more flat in the interior regions.
The normalization
of the SRCUT model fits is fixed at the calculated value of 0.024 Jy at 1 GHz as mentioned earlier. 
Thus, it is clear that there are three components in the spectrum of
the X-ray nebula of GK Per: A cool thermal component associated with the
forward shock/transition zone that shows line emission; a deeply embedded
hot thermal component without emission lines 
and a "localized" embedded non-thermal power law (i.e., Synchrotron) 
emission component coming from the
region where the shell interacts with the circumbinary medium around the nova.
An elaboration of this scenario is in section [7.1].
 
\section{THE Ne IX EMISSION LINE IMAGE AND DETAILED IMAGING OF THE REMNANT}

The nebula displays a different structure compared with the broad band
 \cha image at the emission line energy
detected around 0.907 keV.
The Ne IX emission line image shows a symmetric cone-like shape
centered at the SW direction (see Figure 8) and the  flattening on the SW is more 
prominent. The image is prepared by extracting the
energy channels (0.8-1.0 keV)
forming the full width of the line at the continuum level and subtracting 
on pixel-to-pixel basis
a suitable continuum image from the line image using PI channels corresponding to 1.12-1.4 keV.
This continuum image, also, subtracts out any central source photons since it includes
both the source and the shell continuum in the channel range. I need to note that
some of the line emission and diffuse neon emission 
could have been lost in this process, but Figure 8 is a 
cleaned image.
Both of the line and continuum images are regrouped by two pixels and 
smoothed by a Gaussian of $\sigma$=2$^{\prime\prime}$ before subtraction.
The resolution of the final image (continuum subtracted) 
is 1$^{\prime\prime}$ per pixel and shows emission
from the shell that is 2$\sigma$ above the background.
The neon image of the shell
clearly reveals a cone, a one sided shell (i.e., hemisphere), where the remnant rams into a 
wall like structure
and is compressed in that direction. The image also shows a very lumpy
morphology with clump sizes limited by the resolution of the image
$\sim$ 2$^{\prime\prime}$. The size of the cut-cone is $\sim$
63$^{\prime\prime}$ in the flattened zone and $\sim$ 112$^{\prime\prime}$
-116$^{\prime\prime}$ in the wings toward the North and East in the image.
Assuming 99 years for the elapsed time and a linear expansion law, the
velocities required to create the wings are $\sim$ 2600-2800
km s$^{-1}$ and the flattened zone is $\sim$ 1100 km s$^{-1}$. 
The high velocity necessary for creation of the "wings" detected in the
Ne IX image are more than twice the expansion speed of the nova shell detected
in the other wavelengths and about the
same in the SW region which is the flattened zone. This shows that
the ejecta material is mixed in the forward shock.  
Using the extensions above and the
law of cosines, the cone angle is
calculated to be $\sim$ 74$^{\circ}$. The axis of the cone is in the same direction with the 
axis of the polar cones attributed from the material lying in the 
light-echo region (Seaquist et al. 1989). 

The reverse shock regions
are prone to formation of knots because of Rayleigh-Taylor instabilities.
The width of the interaction region between the forward and reverse shocks
shrinks considerably with increasing compression ratio (due to existence
of nonlinear diffusive particle acceleration) and
the convective instabilities can reach all the way into the forward shock.
As a result, clumps and filaments can be found in the vicinity of 
the forward shock region.
For low compression ratios, any enhancement of the radial magnetic field
could play the same role and
Rayleigh-Taylor instabilities can still extend all the way into the shock front
(see also Blondin $\&$ Ellison 2001 for a discussion). 
This explains the existence of knots of emission (neon knots)
in the forward shock zone of GK Per and also mixing of the ejecta material into the
forward shock. However, this does not
necessarily account for the large asymmetry detected in the remnant.

Therefore, a contributing factor to the above argument 
for the structure in the emission line
would be the existence of undecelerated, cool and dense neon knots in the shell
that have caught up with the forward shock and surpassed it in 
SE and NW directions where the interstellar density is low.
Such a phenomenon was also suggested for the knots in Cas A, Vela, Cygnus Loop and
others. The SE-NW direction is also the site for the reflection nebula, the
symmetric far-IR emission nebulosity, and suggested to be a region
of low density (Lawrence et al. 1995; Tweedy 1995; Bode et al. 1987;
Seaquist et al. 1989). This is also supported by the CO map of the
vicinity of the nova which shows  a relatively empty environment
toward the north and east of the nova and a prominent CO emission
toward the south (Hessman 1989; Scott et al. 1994).
Moreover, detected neon clumps could be formed 
at an early stage in the eruption of the nova and matter could be ejected
as knots in high velocity winds during the early outburst stage (Shore      
et al. 1997). This also implies a binary orientation origin for the 
asymmetry in the remnant together with the non-uniform density profile
of the circumstellar medium around the nova.

In addition to the Ne IX emission line image,  other possible emission line images
were constructed at around 0.58 keV (O VII line) and 0.43 keV (N VI line).
The resulting images indicate that the shape of the nebula at these energies
do not indicate a special structure and the wings seen in the neon image are not detected.
Further energy dependence within  the X-ray nebula is investigated by constructing images 
at different energy bands. Figure 9 displays the image of the X-ray nebula
in three different energy ranges: (a) 0.25-0.50 keV, (b) 0.51-1.50 keV, and (c) 1.51-8.0 keV. 
The first two images show emission from the X-ray nebula that is 2$\sigma$ above the background  
at 1$^{\prime\prime}$ per pixel resolution,
smoothed using a Gaussian with a variable width $\sigma$=1$^{\prime\prime}$-2$^{\prime\prime}$.
They are also corrected for the contamination caused by the ACIS transfer streak and the PSF wings
as described in sec. (4.1). All three bands indicate different structure. The first two bands (up to 1.5 keV; Figure 9a,b)
show generally a filled semi-circular structure brighter on the western side with 
the brightest zone at the flattened rim which is more evident in the second band. This 
resembles the structure of the total band image (see Figure 1) except for the "wing" structure detected
predominantly in the neon emission line (seen also in the second band image). The second band image
is more cone-like owing largely to the motion of the neon knots as described in the previous paragraphs.
The third band (hardest X-rays) image is partially resolved (detected spectrally in sec. 4).
The original discovery of the X-ray nebula of GK Per was also at a resolution of low significance 
with the ROSAT HRI between 0.2-2.4 keV (see Figure 1 in Balman $\&$ \"Ogelman 1999). 
The Chandra observation has recovered most of the emission and structure
missed out by the ROSAT HRI, however, the emission above 2 keV is still not completely resolved, but resembles
the structure of the ROSAT HRI detection and, also, Figure 9a.  
Figure 9d is constructed to stress the resolved emission regions at and above 
3$\sigma$ confidence level
with a lower resolution of 3$^{\prime\prime}$ per pixel. 
It is smoothed using a Gaussian with a width of
$\sigma$=6$^{\prime\prime}$. 
The most prominent region revealed in Figure 9d in the 1.51-8.0 keV band is ellipsoidal with a major axis of 55$^{\prime\prime}$
and a minor axis of 27$^{\prime\prime}$ coincident with the brightest region of the shell in 
0.2-1.5 keV range as shown by the overlay of contours in Figure 9d. Part of this elliptical emission zone is also the region where the Peak Spectrum
is derived and the non-thermal X-ray emission is detected. 
The different structures in the three different energy 
bands support that the X-ray emission from the shell of GK Per has different components. 

\subsection{Comparisons with other Wavelengths}
 
In order to investigate the conditions within the shell and determine the
evolutionary stage of the  remnant, spectral and spatial comparisons
in different wavelengths are necessary. 
Detailed results from the recent data on
GK Per are not available. 
However, a preliminary comparison of the 
X-ray image has been made by {\it aligning} the image  
with the recent images obtained in the optical and radio wavelengths.
Figure 10a shows the X-ray image and an  overlay of contours derived from the
HST image obtained with the [NII] filter (brightest emission line in the
optical wavelengths) (HST image: Shara 2002, private communication).
The combined image reveals that the
[NII] knots and filaments are in general coincident with the central 
circular region of
the X-ray nebula, not with the "wings" detected predominantly in the Ne IX image 
(see also Figure 9). 
Both images do not show a well developed shell and are
brighter on the Western hemisphere rather than the East.
This is not an effect of the X-ray absorption since the spectral fits do not yield
significant variation in the N$_H$ parameter in different directions
within the nebula.
In general, the [NII] image is larger toward the SW of the nova which 
indicates cooling in the forward shock and that the remnant has evolved
in comparison with the aligned images in Seaquist et al. (1989).
The optical shell, mostly composed of knots and filaments, suggests large-scale
density gradients in the vicinity of the classical nova. Though the contours  
look circular in geometry, the HST data indicates an
elliptical velocity gradient in
the shell with the fastest material lying in the NW to SE direction
(Shara 2002, private communication), which is the
direction the elongation in the X-ray nebula lies, as detected
in the image of the Ne IX line and Figure 9b.

Since there is X-ray emission both from the
reverse and forward shocks, the line emission in the UV and optical wavelengths
coming from within the shell and the pre-shock zone
should be affected by the ionization due to X-ray irradiation 
(i.e., particularly, 
the luminosity and evolution of the H${\alpha}$ emission line; 
Chevalier $\&$ Fransson 1994; Chevalier $\&$ Oishi 2003). 
The [OIII] emission line is a characteristics  of the 
shock zones in circumstellar interaction, however the [OIII] emission 
from the shell is weak and mostly suppressed in the SW region where the 
X-ray emission is the brightest (Seaquist et al. 1989; Slavin et al. 1995). 
This could be because the expanding shell is stopped very
efficiently and the shock speeds are low in this region
(shifting the emission to NII and H${\alpha}$ lines). 
The densities of the optical knots in the SW 
are also detected to be high in comparison with the ones in the NE, SE and NW, thus
the prominent emission lines are that of [NII] and H${\alpha}$ (Seaquist et al. 1989). 

The image of the X-ray nebula has been aligned also with a recent VLA image obtained
at  1.425 GHz in 1997 (VLA image: Seaquist 2002, private communication).
Figure 10b shows the X-ray image of the shell with an overlay of radio contours.
The radio image is  mostly circular in structure coinciding
with the central circular part of the X-ray nebula and is slightly smaller than 
the X-ray image (as compared at a level of 2$\sigma$ emission, see also Balman 2002a).
The flattening on the SW is evident in both images. 
An intriguing result of this combination image is that the peak
of the non-thermal radio emission falls slightly offset toward inside of the shell
from the brightest X-ray region (X-ray peak of the nova shell). This indicates
that the particle acceleration
could be occurring at the reverse shock zone consistent with the
high absorption associated with the non-thermal X-ray  component 
(see section 5.0). In addition, the outer
ridge of the radio shell falls in the X-ray bright zone, as well.
An example similar to this has been  detected in the  
\cha observation of  the young 
supernova remnant 1E 0102.2-7219 (Gaetz et al.
2000). Moreover, in Figure 10b the radio wavelength 
contours show a trace of the "wings" at a 1.5$\sigma$ confidence level
(Figure 8, see also the second band image in Figure 9). The part of the radio wavelength (1.425 GHz) image
associated with the "wings" seems to be a part of a much larger and fainter emission which
may not be associated directly with the shell, but the circumstellar/interstellar environment
of the nova. It extends from NW to SE which is in accordance with the 
lower circumstellar density region and the location of the PN and/or the reflection  nebulosity
mentioned earlier. The content of the 
radio spectrum along the "neon wings" is at this time unknown.

\section{DISCUSSION}
\subsection{On the Origin of the X-ray Nebula}

In general, the \cha observation of the shell of nova Persei (1901)
has revealed that the remnant evolves similar to the young supernova remnants
expanding into its circumstellar medium, a leftover wind ejecta expelled
in the course of the binary evolution, and a CO enhanced interstellar medium in the vicinity
(similar to Type II SNRs: Chevalier $\&$ Fransson 1994 hereafter CF94;
see also Truelove and Mckee 1999 for a general review).  
A conspicuous interpretation of the spectrum of the nova shell
is such that the origin of the emission below 2 keV is partly of 
the ejecta (e.g., Ne and N emission lines) and the 
circumstellar matter, namely the forward shock (also the transition zone), 
whereas  the emission 
above 2 keV is of the reverse shock region (owing to the absorption by 
high equivalent neutral column density).  

There is significant difference
(3$\sigma$) between the neutral Hydrogen column densities of the two
components of the nova shell spectrum. This could be explained in the context of SNR
evolution where such a cool layer may exist
between the forward and the reverse shocks in the early phases.
The size ($\Delta$r) of this cold shell is less than 
5$\times 10^{12}$ cm
(less than 0.0005 arc seconds at 470 pc distance) 
for an SNR 100 years old (CF94),
with a range of density  
2$\times 10^{8}$ to 3$\times 10^{11}$ cm$^{-3}$  (CF94). 
Multiplying the density range with the possible size of this region
yields a value of N$_H$ in the cold shell in a 
range 1$\times 10^{21}$-2$\times 10^{24}$ cm$^{-2}$. 
The column density in the cool layer decreases with time and can be expressed
as N$_{cool}$$\simeq$9$\times 10^{22}$(n-4)(t/11.57 days)$^{-1}$ cm$^{-2}$
(FLC96). 
I have assumed 1200 km s$^{-1}$
for the expansion speed, a power law index of 2 for the circumstellar density
gradient, an $\dot{M}$=1$\times 10^{-5}$M$_{\odot}$ yr$^{-1}$ for the
circumstellar wind and a wind velocity of 10 km s$^{-1}$. For a
range of n from 10 to 200 (n=power law index for the density gradient 
in the ejecta), 99 years of age
yields N$_{H}$$\simeq$1.2$\times 10^{20}$-8.7$\times 10^{21}$ cm$^{-2}$ in the cool layer.
SNRs at the age of 100 yrs are about 2 pc in radius
(4 pc in size) whereas the radius of GK per is only 0.11 pc.
The size of an SNR at the same evolutionary age with GK Per is larger by a factor of 20 at the
least. Thus, time may  disperse a cool shell between the forward and reverse shocks in SNRs 
of high velocities and large sizes,
but, since GK Per is considerably slower (compared with an SNR 100 years old)
and more compact with respect to an SNR at the same evolutionary age, 
this cool absorbing layer could stay intact. 
Moreover, the size of GK Per is equivalent to an SNR a few years old.
In such cases, a distinct cool layer is likely 
(if the reverse shock is radiative) and the emission from
the reverse shock would be absorbed.

The high X-ray absorption detected for the shell of GK Per between 0.5-2 keV is not 
largely due to the neutral hydrogen, but to the neutral metals owing to the fact that
the absorbing layer is the cooled part of the reverse shock/ejecta itself. The composition of the 
ejecta is enhanced in metal abundances of N, Ne and possibly O. Even the 
circumstellar wind material is at least of solar composition. 
The absorptivity per hydrogen atom greatly increases above
0.1 keV due to the existence of metals like N, O, Ne, and Fe 
of solar abundances (Wilms, Allen $\&$ McCray 2000 and references therein) 
which would yield large equivalent column densities of neutral hydrogen.
The photoelectric absorption cross-section ${\sigma_K}$ $\propto$ Z$^5$ (Z=atomic number)
and the optical depth increases since $\tau$ $\propto$ ${\sigma_K}$N$_{H}$. 
In order to check this, the same composite models in the Tables 1, 2 and 3 are fitted
using the multiplicative models VARABS and VPHABS (Balucinska-Church $\&$ 
McCammon 1992) in XSPEC to model the
absorption in the second component. These models allow calculation of 
equivalent column densities of neutral Hydrogen together with elemental abundances of the 
absorbing medium. The results show that the X-ray data is consistent
with a range of equivalent neutral Hydrogen
column density of 5.0$^{+19.0}_{-2.9}$$\times 10^{22}$ cm$^{-2}$ attained by, in general, a plasma of
solar abundances with  N/N$_{\odot}$=10.9$^{+15.0}_{-9.1}$ and  
Ne/Ne$_{\odot}$= 17.9$^{+9.1}_{-6.0}$. The elemental abundances determined from the cool layer 
are slightly higher (esp. N) than what is recovered from the analysis of the first component
which is the forward shock mixed with the ejecta. Assuming that the cool layer portrays a better
representation of the ejecta, then the elemental abundances of Ne and N could be higher than displayed
on Tables 1 and 3.
The total column density of metals causing the absorption in the cool shell 
is estimated as N$_Z$$<$ 2$\times 10^{20}$ cm$^{-2}$ which yields a total mass in the
cool layer of M$_{cool shell}$$<$1$\times 10^{-6}$M$_{\odot}$ (less than 1.5$\%$ of the ejecta mass).

The reverse shock speed (using the temperature derived for the second component) is found as
872 km s$^{-1}$  and the temperature of the first 
component (forward shock) implies shock velocities of 
400 km s$^{-1}$ for the forward shock calculated using the general relation
$kT_s=(3/16)\mu m_H (v_s)^2$, assuming Rankine-Hugoniot jump conditions
in the absence of cosmic-ray acceleration.  
The average expansion speed detected for the nova is about 1200 km s$^{-1}$
in the optical and UV wavelengths (see also sec 1). This value is consistent
with the values obtained by measuring the size of the X-ray nebula
in the SW direction (see section 6.0), but inconsistent with the speeds calculated for NW and SE
directions (see section 6.0).  The shock speed calculated using Rankine-Hugoniot 
jump conditions for the first 
component associated with the forward shock is also not in accordance with the
expansion speed of the shell, at all.

The reverse and forward (circumstellar) shock temperatures scale with 
T$_{rev}$ = ($(3-s)^2$/$(n-3)^2$) T$_{cs}$, where {\it s} is the power law index of the
circumstellar density gradient and {\it n} is the index for the 
radial dependence in 
the ejected material as noted earlier (FLC96). 
Following this, the ratio of
the forward and reverse shock temperatures for the nova remnant are about 
4-16 and the shock 
velocities are about 2-4 (i.e. s=1-2 and n=7). As discussed in the
previous paragraph, n is larger than 8 (for a radiative reverse shock) 
and thus the actual contrast
between the forward and reverse shock temperatures are expected to be
larger than the above value.
This approximation yields a minimum
estimation of about  1600-3200 km s$^{-1}$ for the forward shock 
speed ($\sim$ 8 keV) using the computed reverse shock velocity.
There is no component detected in the X-ray spectrum
with the stated shock speed range above and the corresponding temperature.
{\it Therefore, the X-ray temperatures does not follow from the shock speeds/expansion speeds
as expected from simple shock evolution.} 
Some of the detected electron temperatures for the blast wave 
interaction region in 
young SNRs are found to be as low as $\sim$ 10$^{6-7}$ K
(e.g., 1E 0102.2-7219: Hughes et al. 2000, Ellison, Slane $\&$ Gaensler 2001;
Kepler: Decourchelle et al. 2000; SN 1006: Bamba et al. 2003).
This range of shock temperatures are at least a factor of 10 less than
what would be expected from such young SNRs.
Recent theoretical calculations show that
the existence of nonlinear diffusive shock acceleration 
(particle acceleration at the shock zones) modifies shocks, changing their
structure and evolution (Decourchelle et al. 2000; Ellison, Decourchelle $\&$ Ballet 2004
and references therein).                    
The particle acceleration at the shock zones (forward or reverse shocks) 
causes the Mach number to decrease in the post-shock (i.e., sub-shock) regions making 
the shocked gas 
more compressible (increasing the compression ratios) resulting in a
transition region between the reverse shock and the forward shock that is 
narrower and denser.  As the 
efficiency of the acceleration increases, the post-shock temperature (esp. for protons) cools 
significantly (e.g., factor of 10; Ellison et al. 2004 and  
references therein). Therefore, existence of particle acceleration modifies
the shocks and Rankine-Hugoniot conditions are no longer valid.
Consequently, the  
low temperature (0.1-0.3 keV) and the inconsistent shock speed (400 km s$^{-1}$)
calculated for the first component of the
entire shell spectrum associated with the forward shock could be
well explained by nonlinear diffusive
shock acceleration. It is important to note that the higher densities
and lower temperatures will also lead to more rapid Coulomb equilibration
than expected in cases without acceleration. 

Considering both of the components and derived electron densities
in sections (4.3) and (4.4), the mass in the
X-ray emitting nebula is estimated as (2.1-38.5)$\times 10^{-4}$
M$_{\odot}$ using the simple relation, 
M$_{shell}$ $\simeq$ n$_e$ m$_H$ V$_{eff}$.
The calculated shocked mass  
is consistent with the ejecta mass expected 
in classical nova explosions (see sec[1]) and the range is
10-50 times the ejecta mass.
The expansion speed for nova Persei 1901 yields a 
total kinetic energy of
(1.1-6.2)$\times 10^{45}$ ergs for the ejecta (1/2 M$_{ej}$v$^2_{ej}$ and for M$_{ej}$ and v$^2_{ej}$ see the
Introduction). 
The total explosion energy of a classical nova event is of the order of 
10$^{46}$-10$^{47}$ ergs (i.e., binding energy of the envelope of the
white dwarf;  E$_0$= 2GM$_{WD}$M$_{env}$/R$_{WD}$). 
Assuming that the shell will be radiative at the time maximum X-ray luminosity is reached,
then L$_{max}$$\simeq$E$_0$/t$_{max}$. If t$_{max}$ were 99 years (time since the eruption),
we can calculate that L$_{max}$$\simeq$10$^{37-36}$ ergs s$^{-1}$.
The kinetic energy dumped in the ejecta (about 1$\%$ of the total 
explosion energy) translates to about L$_{max}$$\simeq$10$^{35}$ ergs s$^{-1}$ for a 
radiative shell. The reverse shock luminosity is expected to
decrease in time by a factor of 100-1000 times after 100 yrs (CF94) and the luminosity
should be, then, in a range 10$^{33-32}$ ergs s$^{-1}$ which is in accordance with the detected
luminosity of the second component (the reverse shock; L$_x$=(6.6-0.14)$\times 10^{33-32}$ ergs s$^{-1}$). 
On the other hand, given the low luminosity of the shell originating in the forward shock  
a few $\times$ 10$^{32}$ ergs s$^{-1}$
and the fact that the calculated maximum shocked mass is larger than 
the ejecta mass,  
the remnant is most likely evolving in the Sedov phase as an adiabatic remnant. 

It is of evolutionary importance to note that the environment of GK Per resembles the class of
SNRs that interact with molecular clouds (Seaquist at al. 1989). 
The Figure 1 and 9  reveals that the shell has a lumpy structure
which could, also, support the existence of cloud crushing; shocks with the molecular clumps
(White $\&$ Long 1991). A part of the emission of the second component could be due to this.
Interactions of SNRs with dense environments have been modeled and old remnants
like W44 and IC443 are recovered  as the candidates for such SNR evolution with a colder radiative shell in the X-rays
and a hot isothermal gas around 1 keV at the center (Rho $\&$ Petre 1998; Shelton et al. 1999; 
Chevalier 1999; Bykov et al. 2000).
The modeling of SNR W44 (interacting with a dense inhomogeneous region
in the vicinity of a molecular cloud) yields
a one sided X-ray emitting hemisphere, a hot interior and a colder radiative
shell as modeled by Cox et al. (1999). This has a remarkable resemblance in {\it structure} 
to the remnant of
GK Per except that the outer shell is not radiative and cold due to age, but rather adiabatic
and cold most likely as an effect of the particle acceleration and modified shock evolution.
Overall, the nova remnant of GK Per could be a younger remnant that will eventually resemble 
older remnants like  W44 or IC443. 

The photon emission in the X-ray wavelengths presumably arising from the accelerated
particles could be a non-thermal bremsstrahlung process resulting from a particular
non-Maxwellian distribution of accelerated electrons (Asvarov et al. 1990) or synchrotron emission
from a power law distribution of such electrons.  
The non-thermal synchrotron 
emission from the nova remnant as reported in section (5.0) is coming from the brightest
X-ray emission region which is  most likely an X-ray filament.
Lately, such non-thermal filaments have been recovered in SNRs with similar photon indexes
in a range 2.1-2.5 (e.g., Bamba et al. 2003; Uchiyama et al. 2003, 
Ueno et al. 2003). 
The spectrum of accelerated particles in the X-ray band should be
highly sensitive to the local neutral
density and have a non-thermal/thermal bremsstrahlung component at high n$_0$ for hard X-ray emission.
The non-thermal bremsstrahlung intensity is 1.0$\times 10^{15}$$n$N$_0$E$^{-\Gamma}$/($\Gamma$-1)
cm$^{-3}$ erg$^{-1}$ s$^{-1}$ (cf. Chevalier 1999).
The emission from radio to the X-ray band is dominated by synchrotron emission from energetic
electrons for B $>$ 10 $\mu$G with bremsstrahlung dominating X-rays for B $\le$ 3$\mu$G
(see discussion in Ellison, Berezhko, $\&$ Baring 2000 and Ellison et al. 2004 for more details). 
The estimations on the circumstellar
magnetic field intensities in the non-thermal radio peak region and the following outer ridge of GK Per 
are 75 and 47 $\mu$G, respectively 
(Seaquist et al. 1989). As a result, one expects a synchrotron X-ray emitting region co-spatial
with the  
peak and outer ridge detected in the radio wavelengths of GK per as presented in this paper. 

The X-ray luminosity and the power law index from the brightest X-ray (peak) 
region can be used to 
determine the non-thermal electron density and the total energy in the accelerated electrons.
The accelerated electron population will have a distribution in basic terms
N$_e$(E)=K\ E$^{-s}$ with an individual electron losing energy via synchrotron process as
\begin{equation}
{-\dot{E}} =  {4 \over 3} \sigma_{T} c \left({E  \over m\ c^{2}} \right)^{2} \left({B^{2} \over 8 \pi}\right),
\end{equation}
averaging over all pitch angles and taking electron velocity $\sim$ c, where $\sigma_{T}$ is the Thomson
cross-section. The electron distribution described above will have a multiplicative 
exponential decay term to explain the age-limited, and necessarily  
the loss-limited or escape-limited SNRs more properly 
(see Reynolds 1998 and references therein). Following from the prescriptions of Reynolds (1998), the ratio of the
maximum electron energy for the remnant of GK Per and the theoretical energy (i.e., E$_f$ in eq. 27 and 
26 of Reynolds (1998)) of an initially infinitely energetic
electron in the same magnetic field (as in the remnant of GK Per) 
radiating for the same duration time t predicts that the remnant is age-limited (young for synchrotron
losses to dominate or particle escape to be prominent). 

The injection efficiency for the remnant
($\eta=n_{non-thermal}/n_{thermal}$) is proportional to the ratio of the electron number densities
of the non-thermal and thermal electrons. I estimate that $\eta$ = 1.4$\times 10^{-3}$ for 
the brightest X-ray (peak) region of the remnant of GK Per.
This is about the same order of magnitude with the $\eta$ reported for SN 1006 (Bamba et al. 2003)
which is the most efficient SNR known to accelerate particles to very high energies close 
to the knee. The non-thermal electron density 
is calculated as n$_{non-thermal}$ = 3.01$\times 10^{-3}$ cm$^{-3}$ for
a filling factor of 1 (n$_{thermal}$=2.0 cm$^{-3}$ for f=1).  
A basic power law electron distribution is assumed for the calculation, along with the integration limits : 
(1) a minimum energy (the injection energy) 
of 1 keV (the temperature of the thermal plasma of the second component), (2) a maximum energy of 15 TeV. 
In accordance with the photon extraction region, 
a volume of 2.8$\times 10^{52}$ cm$^{-3}$ is also, assumed. 

The non-thermal energy density of the  brightest X-ray emission (peak) region is 
estimated as 3.0$\times 10^{-10}$ erg cm$^{-3}$.
This is calculated using the same integration limits  and volume described in the above paragraph together with
a distribution of power emitted by the given power law electron distribution 
(i.e., [dP(E)/dE]$\simeq$$\dot{E}$[dN(E)/dE]).
The corresponding energy densities of the magnetic field and the thermal plasma are 
2.3$\times 10^{-10}$ erg cm$^{-3}$ (u$_B$=B$^2$/8$\pi$ ) and 
5.1$\times 10^{-9}$ erg cm$^{-3}$ (u$_{thermal}$=3/2 n$_e$kT), respectively. 
The energy densities are a factor of 10 higher compared with SN 1006 by Bamba et al. (2003).
The total energy in the
thermal plasma, the magnetic field and the non-thermal electrons are then, 2.2$\times 10^{45}$ ergs,
9.5$\times 10^{43}$ ergs (if total volume is assumed and 6.2$\times 10^{42}$ ergs if the photon 
extraction region is assumed), and 1.2$\times 10^{44}$ ergs (if total volume is assumed and 8.4
$\times 10^{42}$ ergs if the photon extraction region is assumed),
respectively. 
As a result, the magnetic field is almost/in  equipartition with 
the non-thermal electrons (taking that these calculations are estimations). 
The non-thermal protons should carry more
energy than the non-thermal electrons (in a range 1-2000 times 
the total energy of electrons [2000$\sim$m$_p$/m$_e$]). 

There are strong indications of possible nonlinear diffusive particle
acceleration occurring in the forward shock zone as well as the reverse shock zone; 
(1) the unexpectedly cool X-ray temperature, (2) the strongly compressed interaction region
between the forward and reverse shocks where the ejecta material is detected to be mixed in.
However, including a second power law or SRCUT
model of emission in the fits applied on  the X-ray data below 2 keV does not yield  
physical or statistically consistent results unless the first component is fitted with
a bremsstrahlung model and three Gaussians for emission lines along with the 
SRCUT/power law model. Inclusion or
exclusion of the SRCUT/power law model does not affect the reduced ${\chi}^2$ of the fits.
Thus, an upper limit on the non-thermal X-ray flux from the forward shock is F$_x$$<$1.0$\times 10^{-14}$
erg s$^{-1}$ cm$^{-2}$ 
with a photon index of 2.8$^{+4.4}_{-2.5}$ and a luminosity $<$ 3.0$\times 10^{29}$ erg s$^{-1}$. 
The SRCUT model implies a radio spectral index of 0.8$^{+0.2}_{-0.1}$.
There can be several explanations for this: (a) 
the energy going into particle acceleration is less, (b) the synchrotron 
losses and/or particle escape is more effective  compared with the 
reverse shock yielding a lower X-ray flux for the forward shock zone, (c) the forward shock is
under-ionized as discussed in section (4.3) which results in a lower injection efficiency
and thus less particles in the non-thermal regime. 
Ellison et al. (2004) shows that nonlinear diffusive shock acceleration in the forward and reverse shocks
could evolve differently. The emission from the reverse shock may increase out to 1000 years
and the efficiency diminishes after that together with the luminosity. The emission
and efficiency from the forward shock is expected to increase slowly out to 40000 years and may be 
small for young SNRs (younger than 1000 yrs). 
 
\subsection{The Classical Nova Remnants}

The studies of resolved old nova shells provide
an opportunity of investigating several facts of the evolution of novae long
after the initial eruption has subsided.  
Spectroscopy and imaging of the 
nebular remnants of old novae have been a diagnostic of physical conditions in
(mode of excitation) and the morphology of the nova shells.
Optical spectroscopy/imaging of nova shells has
been widely used for this purpose.
By direct imaging method, the ejecta of several old novae have been spatially
resolved with evidence/detection of polar blobs and equatorial rings in many
of them (Slavin et al. 1995; Gill $\&$ O'Brien 1998, 2000; 
Krautter et al. 2002; Bode 2002 and references therein).
The expansion velocities of material in a nova shell are suggested to
vary smoothly as a
function of position angle from a maximum in the polar regions to a minimum at
the equator, which results in a prolate asymmetry (Lloyd et al. 1997).
In addition to this, abundance gradients are expected as a result of TNRs
(Thermo-Nuclear Runaway) in rotating oblate white dwarfs as detected from observations (RR Pic, 
DQ Her, V1370 Aql, V838 Her, etc.; Scott 2000). Recently, the HST imaging of
HR Del indicates several emission knots in [OIII], [NII] and H$\alpha$; particularly
a bipolar structure in the first two compared with the latter (Harman $\&$ O'Brien 2003).
Analysis of HR Del reveals that there are also indications of wind (from the central source)
interaction with the old remnant shell affecting its evolution and structure. 
In general, optical and UV wavelength measurements of electron temperatures
in the nova shells indicate two groups of old remnants;
hot (T$_e$ $>$ 10$^4$ K) and  cold (T$_e$ $<$ 10$^4$ K)
nova shells (Williams 1992, Ferland 1984). Some of the nova shells have been
found to decelerate in time, such as DQ Her, GK Per, V603 Aql, T Pyx, and
V476 Cyg, suggesting the existence of circumstellar interaction
(Duerbeck 1987). 

The first prediction of classical nova remnants as
non-thermal radio sources has been made by Chevalier (1977).
However, there have not been any X-ray or radio wavelength (Bode et. al. 1987)
detections of old classical nova remnants
except for GK Per.
The detailed theoretical framework on the
hydrodynamical shock evolution in classical nova remnants 
is being developed (See Bode 2002 for a review). 
The circumstellar
interaction has been modeled for recurrent novae (RS Oph:
 Bode $\&$ Kahn 1985; T Pyx: Contini $\&$ Prialnik 1997). These models
(esp. for T Pyx) predict the existence of a forward moving blast wave and a
reverse shock moving into the ejecta as a result of colliding shells from the eruptions. 
Their model also predict a cold layer of material between the forward and reverse shocks
(they do not study it in detail).
The models indicate high electron densities of the order of a few hundred 
cm$^{-3}$ in the shock zones. Another intriguing result from these studies is that
the magnetic field is enhanced by factors of 10 to 100 due to instabilities and 
turbulent motion. 

The power from
a simple shocked-shell model of thermal origin, as an X-ray emitting nebula,
can be modeled
assuming a 0.25 filling factor for the volume of the shell (assumed
to be spherical) and n$_e\sim$ 4n$_o$ (from strong shock conditions);
\begin{equation}
 L_T\simeq 3.1\times 10^{33} T_{7}^{0.5} n_o^2  R_{18.5}^{3}.
\end{equation}
The temperature T is in units of 10$^7$ K and the radius of the emitting shell
R is in units of 3.1$\times 10^{18}$ cm. The fast nova ejecta will start to cool as
it sweeps
up an equal amount of circumstellar material at R$_{cool}$ and t$_{cool}$.
Once it sweeps up about 10 times its own mass, it will proceed into the
Sedov phase;
\begin{equation}
R_{cool}\simeq 0.03 (M_{-5}/n_o)^{1/3} pc,  
\end{equation}  
\begin{equation} 
t_{cool}\simeq 30 (M_{-5}/v_{1000}^3 n_o)^{1/3} yrs,
\end{equation}   
\begin{equation}  
t_{Sedov}\simeq 64 (M_{-5}/v_{1000}^3 n_o)^{1/3} yrs. 
\end{equation}  
For n$_o$ $\simeq$ 0.1-10 cm$^{-3}$, kT $\simeq$ 0.1-10 keV
and R $\sim$ R$_{cool}$, the range of X-ray luminosity is a few$\times 10^{34}$
to a few$\times 10^{28}$ ergs s$^{-1}$ as the remnants start to cool. 
Assuming adiabatic remnants (at the onset of cooling),
about 1$\%$ of this radiation should be detected. Thus, the maximum limit
of X-ray emission is then about 10$^{32}$ ergs s$^{-1}$ 
which is in good agreement
with the X-ray luminosity of the remnant of GK Per 
(i.e., brightest remnant). 
The time to start to progress into the Sedov phase is $<$ 96 yrs for GK Per, calculated using
equation (7) which is consistent with the interpretation that the remnant
is in the Sedov phase.  
I would like to note that the X-ray synchrotron emission from the remnant of GK Per
is a special case and a result of the circumstellar environment and the magnetic fields.  

Theory would expect that the remnants should become fully radiative in the later stages of the shock evolution
(i.e. Snowplow phase), increasing the luminosity in excess of $\times 10^{35}$ ergs s$^{-1}$.
However, by that time the remnant will expand to a much larger size where the decreasing surface
brightness will affect this evolution and may no longer permit the remnant to be detected.

\section{SUMMARY AND CONCLUSIONS}

The nova shell of GK Per is the first classical nova remnant resolved
and detected in the X-ray wavelengths. \cha Observatory
has revealed the spectrum and detailed morphology of this remnant for 
the first time
and this has helped to raise/resolve important questions 
related to classical nova eruptions, hydrodynamic
flows and relativistic particle generation. In general, the nova remnant of GK Per
evolves similar to young, type II SNRs and a shell is constructed via the interaction
of the ejecta with the circumbinary medium which is a leftover wind and/or a Planetary Nebula. 
The X-ray nebula around GK per is brightest on
the SW quadrant and toward West with a lumpy morphology. There is a central semi-circular
region and "wing" like extensions toward SE and NW. 
The central circular region resembles the radio/VLA shell and the 
optical/HST images
with several clumps/knots in the X-rays. Some features are collection of clumps and filamentary structures,
also, indicating radial elongations. 

The X-ray shell has a  symmetric cone-shaped structure detected
in a Ne IX emission line with a cone angle of about 74$^{\circ}$ to the central
source which indicates the shock front. The shell is  most likely
expanding into  a less dense medium in the NW and SE directions which
comprises the "wings" with speeds 
(i.e., 2600 km s$^{-1}$) more than 
twice the expansion speed in the SW direction (i.e., 1100 km s$^{-1}$). 
The VLA/radio image of the remnant shows a trace along the "wings" detected
in the X-rays as part of a larger diffuse radio emission stretching
from NW to SE in the direction of the reflection nebulosity.
The neon line emission and the neon wings detected
toward the NW and SE, moving with speeds $\sim$ 2600-2800 km s$^{-1}$, 
originate from the clumped ejecta due to
instabilities in the post-shock region. Such
knots are expected to
transit into the forward shock zone in time with possibly the aid of a circumstellar magnetic field
and compression of the interaction region due to the effects of nonlinear diffusive 
shock acceleration (particle acceleration). It is also possible that some of
these knots could have been created within the remnant
or even at the onset of the nova explosion and have not significantly 
decelerated, eventually surpassing the shock zone particularly in the NW and SE
directions where the interstellar density is low.
The NE portion of the circumstellar environment does not show a distinct
sign of interaction (i.e., a bright rim) as in the SW and lacks
line emission. In general, X-ray nebula is the largest among the detected
radio and optical shells.
The shape of the X-ray nebulosity is a combination of the
characteristics of the
circumstellar medium around the nova and the geometrical projection effects that requires 
a special treatment with hydrodynamical models. Creation of a hemispherical X-ray shell,
as detected for GK Per in this paper, is theoretically expected in regions with high ambient density
gradients as revealed by studies in SNRs evolving in molecular clouds (e.g. W44, Cox et al. 1999).
 
The nova shell of GK per is a site where energetic electrons are produced
by diffusive shock acceleration (particle acceleration) confirmed by the
\cha observation via the discovery of the synchrotron emission from the non-thermal
electrons in the reverse shock zone. The detected power law emission 
in the X-rays reveals the same spectral index as in the radio wavelengths 
recovered with the fits using a SRCUT model of emission.
The detected X-ray photon index is 2.3$^{+1.5}_{-0.9}$, the spectral index is 0.68$^{+0.03}_{-0.15}$ 
and the unabsorbed luminosity is 4.6
$\times 10^{30}$ erg s$^{-1}$. The remnant is age-limited. The non-thermal
electron density is about 3.01$\times 10^{-3}$ (f=1) and the efficiency of 
injection in the reverse shock is about 1.4$\times 10^{-3}$. This value 
is similar to the efficiency derived for SN 1006
which is the most efficient SNR in particle acceleration. However, the derived luminosity for GK Per is lower
because the emitting volume is smaller. The cool X-ray temperature (0.1-0.3 keV) of the 
forward shock indicates that there is nonlinear diffusive shock acceleration occurring in this 
region, also. However, the efficiency and the luminosity is low (L$_x$$<$3$\times 10^{29}$ erg s$^{-1}$
and $\Gamma$$>$2.8) and can not be resolved from the thermal emission. 
This complies with the theoretical expectations that for young SNRs the contribution from particle acceleration  
into the energetics of the systems will be small. 

The calculated break frequency for the remnant shows that GK Per is also a strong candidate for
gamma-ray emission (out to the knee) due to the accelerated particles in the reverse
shock region with E$_{max}$ about 15-30 TeV (this should be more or less correct for the 
forward shock region, too). 

The X-ray nebula has a low temperature component (below 2 keV), the first component,
that is a (non-equilibrium ionization) thermal plasma close to ionization equilibrium, possibly dominated by emission
lines, with kT$\sim$ 0.1-0.3 keV and an X-ray luminosity $\sim$
4.3$\times 10^{32}$ ergs s$^{-1}$. This component shows the  
He-like Ne IX (i.e, most contribution may be coming from a forbidden emission line) and He-like N VI line emission
derived from the spectral fits (and a possible O VII emission line). No H-like line emission is detected.
This component is associated with the forward shock region and the transition zone
owing to the low neutral Hydrogen column density. Thus, it is a mixed plasma
of shocked circumstellar matter and ejecta material. The electron density 
within this region is 0.6-11.2 cm$^{-3}$ (for a filling factor of 1). The n$_0$ in the circumstellar
medium is $\le$60 cm$^{-3}$ (n$_0$$\sim$4.1 cm$^{-3}$). 
 
The higher energy component of the entire shell spectrum above 2 keV is most likely a plasma emission 
of bremsstrahlung origin with no resolved emission lines. The fits using  a thermal bremsstrahlung model of 
emission reveal a shock temperature of
 kT = 1.04$^{+1.7}_{-0.5}$ keV and an X-ray luminosity 
(6.6-0.14)$\times 10^{33}$ erg s$^{-1}$. It is found to be
heavily absorbed with N$_H$ = (4.0-22.0)$\times 10^{22}$ cm$^{-2}$
suggesting that there is a cold layer of material between the two
X-ray components and the second component is most likely originating from the shocked gas in the 
reverse shock zone. The X-ray analysis reveals that this cold layer
is a plasma of solar abundances with N/N$_{\odot}$$\sim$11  and Ne/Ne$_{\odot}$$\sim$18
where the high metalicity of the ejecta in this cooling layer is 
largely responsible for the high optical depth between 0.2 and 2 keV.  
It is also important to note that this high column density of neutrals
absorbs the emission below 1.5 keV where most of the line emission would be detected.

The standard nova theory predicts enrichment of metal abundances
produced prior to the outburst via mixing processes.
This observation reveals for the first time a minimum 
enhancement of neon 13-21  and nitrogen 1-5  times its solar number fraction
from a remnant in the X-rays that is 100 yrs old. Moreover, the existence of 
enhanced neutral abundances of neon and nitrogen
in the "cold absorbing shell" supports this finding.
As a result, this confirms that
the outburst was a result of a violent TNR on an ONeMg WD, which
would, also, explain the existence of a blast wave. The high nitrogen ratio
is a direct evidence of the $\beta^+$ unstable nuclei which must have been
ejected with the blast wave from the nova in the early stages.

The kinetic energy density of the ejecta is (0.3-1.5)$\times 10^{-8}$ erg cm$^{-3}$
where the maximum limit comes from the calculated shocked mass (i.e.  
(2.1-38.5)$\times 10^{-4}$ M$_{\odot}$). 
The energy densities of the thermal plasma, magnetic field and non-thermal electrons 
are 5.1$\times 10^{-9}$ erg cm$^{-3}$, 2.3$\times 10^{-10}$ erg cm$^{-3}$ and 
3.0$\times 10^{-10}$ erg cm$^{-3}$, respectively, implying that the magnetic field and the 
non-thermal electrons are almost/in equipartition at the shock zone.
The energy budget of the nova remnant indicates that not all the kinetic energy
has appeared as thermal energy in the shock zone, in accordance with the 
fact that the remnant is young. Thus, considerable energy is still
going into expansion of the shell and about less than 34$\%$ appears as thermal energy 
with about less than 2$\%$ in the accelerated particles.
The low X-ray luminosity (4.3$\times 10^{32}$ erg s$^{-1}$) of the forward  
shock indicates that the remnant is adiabatic. The reverse shock is radiative. 
Overall, the nova shell of GK Per resemble a Sedov remnant (of SNRs). The cooling in the
forward shock is also detected in the HST [NII] images (very similar to H${\alpha}$ images) aligned with the
X-ray data (see Figure 10a).

The nova shell of GK Per remains to be unique with its characteristics over 
the entire electromagnetic spectrum. It behaves as a miniature SNR with
typical characteristics, yet it is the first detection of CNRs (Classical Nova 
Remnants), which can be a new class of X-ray emitters that have considerably low
surface brightness/luminosity and shocked mass 
compared with SNRs; as a result evolving faster
than, but similar to SNRs. The luminosity range for such objects are
between a few$\times 10^{32}$
to a few$\times 10^{26}$ ergs s$^{-1}$ as the adiabatic 
remnants start to cool. After the outburst of a classical nova, the typical cooling
timescale for the hard X-ray emission components  due to wind-wind interactions
is about three years (Balman et al. 1998). However, as the shell 
interacts with its circumstellar environment, it should 
produce about the range of luminosity above depending
on how strong the shock will be (i.e., density contrast/compression ratio) 
and how the expansion will affect the surface brightness of the 
remnant.     
Two other candidates, DQ Her and RR Pic indicate residual emission
coming from the location of their shells (Mukai, Still, $\&$ Ringwald 2003; 
Balman 2002b). Recovering CNRs require simultaneous high sensitivity and 
spatial resolution and pose a challenge given the characteristics of the present
and upcoming X-ray missions. The ACIS detector on
\cha and Epic PN on Newton-XMM 
could discover some of the close-by and luminous CNRs.

\acknowledgements
{The Author thanks E. Seaquist and M. Shara for 
making the VLA and HST
images available before publication. Many thanks, also, to
an anonymous referee for the critical reading of the manuscript.
I would like to also thank M. Bode, D. Ellison, M. Hernanz,
J. Krautter, K. Mukai, H. \"Ogelman, 
E. Schlegel, M. Shara, N. Schultz and P. Plucinsky 
for helpful remarks or help with data handling. This work has made
use of \cha archives and  \cha workshop (2001)/user help desk on several occasions.
Finally, I would like to acknowledge the support from the Turkish Academy of Sciences with
the TUBA-GEBIP (distinguished young scientist award) Fellowship. }

\begin{table}
\label{1}
\caption{Spectral Parameters of the Entire Spectrum of the Nova Shell
for the {\bf Softer X-ray Component in the energy range 0.3-2 keV}
(ranges correspond to 2$\sigma$\ errors)}
{\footnotesize
                                                                    
\begin{tabular}{l|l|l|l} \hline\hline
\multicolumn{1}{l}{}  &
\multicolumn{1}{l}{\ VPSHOCK \tablenotemark{\S{1}}} &
\multicolumn{1}{l}{\ VMEKAL \tablenotemark{\S{2}}} &
\multicolumn{1}{l}{\ BREMSS+3(GAUSSIAN) \tablenotemark{\S{3}}} \\ 
\hline
 N$_H$ ($\times 10^{22}$ cm$^{-2}$) &  0.30$^{+0.07}_{-0.07}$ &  
0.23$^{+0.05}_{-0.04}$ & 0.19$^{+0.18}_{-0.16}$   \\
kT$_{s}$ \tablenotemark{a}  (keV) &  0.11$^{+0.04}_{-0.02}$ & 
0.10$^{+0.02}_{-0.01}$ & 0.27$^{+0.78}_{-0.16}$  \\
Nitrogen  (N/N$_{\odot}$) & 3.1$^{+1.6}_{-2.1}$ & 3.0$^{+0.8}_{-1.2}$ & N/A \\
Oxygen  (O/O$_{\odot}$) & 1.8$^{+1.7}_{-0.8}$ & N/A  & N/A \\
 Neon    (Ne/Ne$_{\odot}$) & 16.3$^{+4.6}_{-3.4}$ & 9.23$^{+6.0}_{-1.2}$ 
& N/A \\
${\tau}$ ($\times 10^{11}$ s cm$^{-3}$) & 0.3$^{+1.4}_{-0.2}$ & N/A & N/A \\
EM \tablenotemark{b,} \tablenotemark{c} ($\times 10^{53}$ cm$^{-3}$) & 
138.6$^{+408.5}_{-111.2}$ & 66.7$^{+133.3}_{-63.24}$ &
52.1$^{+238.4}_{-51.9}$ \\
Line Energy (keV) & N/A & N/A & Bremss: N/A \\
   &   &     &  G1:  0.423$^{+0.010}_{-0.011}$ \\
   &   &     &  G2:  0.557$^{+0.005}_{-0.008}$ \\
   &   &     &  G3: 0.907$^{+0.009}_{-0.007}$ \\
Line Sigma (keV)  & N/A & N/A & Bremss: N/A \\
   &   &     & G1: 0.005$^{+0.010}_{-0.004}$ \\
   &   &     & G2:  0.028$^{+0.008}_{-0.010}$ \\
   &   &     & G3:  0.003$^{+0.015}_{-0.002}$ \\ 
 Flux \tablenotemark{d} & 16.1$^{+118.3}_{-15.0}$ & 3.9$^{+9.1}_{-2.2}$  & Bremss: 0.32$^{+54.0}_{-0.03}$  \\
 &    &  & G1:  0.32$^{+109.0}_{-0.28}$ \\
 &    &  & G2:  0.38$^{+18.32}_{-0.29}$ \\
 &    &  & G3:  0.032$^{+10.9}_{-0.015}$ \\
 $\chi^2_{\nu}$ & 1.3 (84) & 1.5 (85) & 1.3 (87)  \\                      
\hline                                                                          
\end{tabular}                                   
}                                                                      
\tablenotetext{\S{1}\S{2}\S{3}} {The spectral models are fitted 
simultaneously with the models labeled by the same number in Table 2.}
\tablenotetext{a} { 1 keV $\simeq$ 1.2 $\times 10^7 K$.} 
\tablenotetext{b} { Calculated using the normalization constant of the
VMEKAL/VPSHOCK thermal plasma models (see sec[4.1]). A =
(10$^{-14}$/4$\pi$D$^2$)$\times$EM where EM (Emission Measure)
= $\int n_e\ n_H\ dV$ (integration is over the emitting
volume V).}
\tablenotetext{c} {Calculated using the normalization constant of the
Thermal Bremsstrahlung model. \\ A =
(3.02$\times$10$^{-15}$/4$\pi$D$^2$)$\times$E.M.
where E.M. (Emission Measure) = $\int n_e\ n_I\ dV$
(integration is over the emitting volume V).}          
\tablenotetext{d} { Unabsorbed Hard X-ray flux between 0.3 and 10 keV in 
units of
$\times 10^{-12}$ erg cm$^{-2}$ s$^{-1}$. The unconstrained parameters
are kept at their best fit values when calculating model fluxes.}
\end{table}

\begin{rotate}
\begin{table}
\label{2}
\caption{Spectral Parameters of The Entire Spectrum of the Nova Shell for the
{\bf Harder X-ray Component} above 2 keV
(ranges correspond to 2$\sigma$\ confidence level errors)}
{\footnotesize 
 
\begin{tabular}{l|l|l|l|l|l|l} \hline\hline
\multicolumn{1}{l|}{\  Model}  &
\multicolumn{1}{l|}{\ N$_H$ } &
\multicolumn{1}{l|}{\ kT$_{s}$ \tablenotemark{a}} &
\multicolumn{1}{l|}{\ ${\tau}$ } & 
\multicolumn{1}{l|}{\ E.M. \tablenotemark{b} \tablenotemark{c}} &
\multicolumn{1}{l|}{\ Flux \tablenotemark{d}} &
\multicolumn{1}{l}{\ $\chi^2_{\nu}$} \tablenotemark{e} \\
  & ($\times 10^{22}$ cm$^{-2}$) & (keV)
 &  ($\times 10^{11}$ s cm $^{-3}$) & 
($\times 10^{55}$ cm$^{-3}$) & & \\
\hline
NEI$^{\S{1}}$ & 14.2$^{+5.8}_{-11.2}$ & 0.9$^{+0.9}_{-0.3}$ & 0.09$^{+0.41}_{-0.089}$ & 
3.02$^{+13.0}_{-2.5}$ &  27.2$^{+322.8}_{-25.8}$ & ... \\
PSHOCK$^{\S{2}}$  & 13.7$^{+7.3}_{-9.7}$ & 0.96$^{+1.04}_{-0.4}$ & 
0.0-0.13$^{+0.27}_{<}$ \tablenotemark{f} & 2.2$^{+11.1}_{-1.9}$ & 
150.0$^{+50.0}_{-149.6}$ &... \\
BREMSS$^{\S{3}}$ &  12.9$^{+9.1}_{-6.9}$ &  1.04$^{+1.66}_{-0.5}$ & N/A & 
2.4$^{+6.4}_{-2.34}$ & 5.2$^{+29.8}_{-4.45}$ & ... \\
NEI$^{\S{1}}$ & 0.2$^{+0.1}_{-0.1}$ \tablenotemark{g} & 60.0$^{<}_{-51.0}$ & 6.7$^{<}_{-4.3}$ & 
0.001$^{+0.001}_{-0.0005}$ & 0.08$^{+0.06}_{-0.03}$ & 2.4 \\
\hline
\hline
\multicolumn{1}{l|}{\  Model}  &
\multicolumn{1}{l|}{\ N$_H$ } &
\multicolumn{1}{l|}{\ $\Gamma$ } &
\multicolumn{1}{l|}{\ $\alpha$ } &
\multicolumn{1}{l|}{\ Break Frequency } &
\multicolumn{1}{l|}{\ Flux \tablenotemark{d} } &
\multicolumn{1}{l}{\ $\chi^2_{\nu}$} \\
  & ($\times 10^{22}$ cm$^{-2}$) &
    &  & ($\times 10^{18}$ Hz) 
&  & \\
\hline
POWER$^{\S{3}}$ & 16.9$^{+8.1}_{-5.9}$ & 5.9$^{+1.2}_{-2.1}$ & N/A & N/A & 14000.0$^{+286000.0}_{-13000.8}$ 
&  ... \\
SRCUT$^{\S{3} \dagger}$ & 10.5$^{+5.5}_{-4.5}$ & N/A & 0.32$^{+0.25}_{<}$ & 0.05$^{+0.8}_{-0.03}$ & 6.2$^{+32.0}_{-5.44}$ & ... \\
POWER$^{\S{1}}$ & 0.29$^{+0.04}_{-0.04}$ \tablenotemark{g} & 0.04$^{+0.6}_{-0.9}$ & N/A & N/A & 0.3$^{+4.0}_{-0.27}$ & 1.94 \\
\hline

\end{tabular}
 
}
\tablenotetext{\S{1}\S{2}\S{3}} {The spectral models are fitted
simultaneously with the models labeled by the same number in Table 1. $\dagger$
The normalization of the SRCUT model (the radio flux at 1 GHz) is frozen at 0.024 Jy
at the measured value (Reynolds $\&$ Chevalier 1984; Biermann et al. 1995)}  
\tablenotetext{a} { 1 keV $\simeq$ 1.2 $\times 10^7 K$.} 
\tablenotetext{b} { Calculated using the normalization constant of the
VMEKAL/PSHOCK model. A =
(10$^{-14}$/4$\pi$D$^2$)$\times$E.M. where E.M. (Emission Measure)
= $\int n_e\ n_H\ dV$ (integration is over the emitting
volume V).}
\tablenotetext{c} {Calculated using the normalization constant of the
Thermal Bremsstrahlung model. \\ A = 
(3.02$\times$10$^{-15}$/4$\pi$D$^2$)$\times$E.M. 
where E.M. (Emission Measure) = $\int n_e\ n_I\ dV$ 
(integration is over the emitting volume V).} 
\tablenotetext{d} { Unabsorbed Hard X-ray flux between 0.3 and 10 keV
in units of $\times 10^{-12}$ erg cm$^{-2}$ s$^{-1}$.}
\tablenotetext{e} {The values  for $\chi^2_{\nu}$ are the same as in Table 1 
since the two emission models are fitted simultaneously.  Thus, they 
will be omitted from this table.}
\tablenotetext{f} { The range is derived using the values for the two 
parameters $\tau_{up}$ and $\tau_{low}$. $\tau_{low}$ is fixed at the value of
0.0 for convenience, since it is the lower limit for the integration over the
ionized volume and $\tau_{up}$ shows the maximum limit on the ionization
parameter at 2$\sigma$ confidence level.}
\tablenotetext{g} {This value of N$_H$ is derived from the fits with a single absorption parameter
used for both X-ray spectral components.}
\end{table}

\end{rotate}

\begin{rotate}
\begin{table}
\label{3}
\caption{Spectral Parameters of The Brightest Emission Region of the
Nova Shell (ranges correspond to 2$\sigma$\ confidence level errors)}
{\scriptsize
\begin{tabular}{l|l|l|l|l|l|l|l|l} \hline\hline \multicolumn{1}{l|}{\ Model for}  &
\multicolumn{1}{l|}{\ N$_H$ } &
\multicolumn{1}{l|}{\ kT$_{s}$ \tablenotemark{a}} &
\multicolumn{1}{l|}{\ ${\tau}$ } &
\multicolumn{1}{l|}{\ Ne/Ne$_{\odot}$ } &
\multicolumn{1}{l|}{\ N/N$_{\odot}$ } &
\multicolumn{1}{l|}{\ E.M. \tablenotemark{b}} &
\multicolumn{1}{l|}{\ Flux \tablenotemark{c}} &
\multicolumn{1}{l}{\ $\chi^2_{\nu}$}  \\
Component1  & ($\times 10^{22}$ cm$^{-2}$) & (keV)
 &  ($\times 10^{11}$ s cm $^{-3}$) & & &
($\times 10^{53}$ cm$^{-3}$) & & \\
\hline         
VPSHOCK$^{\S{1}}$  & 0.37$^{+0.05}_{-0.04}$ \tablenotemark{d} & 0.1$^{+0.04}_{<}$ &
0.0-2.8$^{<}_{-2.4}$ \tablenotemark{e} & 7.6$^{+13.0}_{-6.6}$ & 1.9$^{+2.2}_{-0.9}$ &
0.7$^{+0.4}_{-0.2}$ & 0.2$^{+0.7}_{-0.15}$ &...\tablenotemark{f} \\
VMEKAL$^{\S{2}}$  &  0.25$^{+0.08}_{-0.05}$ \tablenotemark{d} & 0.1$^{+0.02}_{-0.01}$ &
N/A  & 7.3$^{+7.7}_{-4.1}$ & 2.8$^{+5.2}_{-1.8}$ & 11.7$^{+15.0}_{-7.2}$ & 2.9$^{+12.7}_{-2.2}$ & ...\tablenotemark{f} \\
\hline
\hline
\multicolumn{1}{l|}{\  Model for}  &
\multicolumn{1}{l|}{\ N$_H$ } &
\multicolumn{1}{l|}{\ $\Gamma$ } &
\multicolumn{1}{l|}{\ $\alpha$ } &
\multicolumn{1}{l|}{\ Radio Flux } &
\multicolumn{1}{l|}{\ Break Frq. } &
\multicolumn{1}{l|}{\ Flux \tablenotemark{c} } &
\multicolumn{1}{l}{\ $\chi^2_{\nu}$} \\
Component2  & ($\times 10^{22}$ cm$^{-2}$) &
 &   & at 1 GHz (Jy) & ($\times 10^{18}$ Hz)
 &  & \\
\hline         
POWER$^{\S{2}}$ & 4.9$^{+5.1}_{-3.7}$ \tablenotemark{d} & 2.3$^{+1.5}_{-0.9}$ & N/A & N/A & N/A & 1.7$^{+73.0}_{-1.4}$ & 1.15 \\
SRCUT$^{\S{1}}$ & 5.0$^{+5.5}_{-1.7}$ \tablenotemark{d} & N/A & 0.68$^{+0.03}_{-0.15}$ & 0.024  & 1.2$^{+2.3}_{-0.3}$ &
 1.7$^{+125.3}_{-1.6}$ & 1.16 \\
\hline
POWER$^{\S{2}}$ & 0.38$^{+0.17}_{-0.23}$ \tablenotemark{g} & -0.2$^{+0.7}_{-0.8}$ & N/A  & N/A & N/A & 30.0$^{+55.0}_{-18.0}$ &  1.9 \\
SRCUT$^{\S{1}}$ & 0.40$^{+0.06}_{-0.15}$ \tablenotemark{g} & N/A & 0.87$^{+0.06}_{-0.02}$ & 0.024 & 4.5$<$  & 0.03$^{+0.02}_{-0.01}$ & 3.0 \\
\hline
\hline
\multicolumn{1}{l|}{\ Model for } &
\multicolumn{1}{l|}{\ N$_H$ } &
\multicolumn{1}{l|}{\ kT$_{s}$ \tablenotemark{a}} &
\multicolumn{1}{l|}{\ ${\tau}$ } &
\multicolumn{1}{l|}{\ E.M. \tablenotemark{b}} &
\multicolumn{1}{l|}{\ Flux \tablenotemark{c}} &
\multicolumn{1}{l}{\ $\chi^2_{\nu}$} \\       
Component2  & ($\times 10^{22}$ cm$^{-2}$) & (keV)
 &  ($\times 10^{11}$ s cm $^{-3}$) &
($\times 10^{53}$ cm$^{-3}$) & & \\
\hline
PSHOCK$^{\S{2}}$ & 4.2$^{+5.0}_{-1.8}$ \tablenotemark{d} & 20.9$^{<}_{-17.0}$ &
96$^{<}_{-95.9}$ & 1.1$^{+14.6}_{-0.5}$ & 0.8$^{+10.7}_{-0.4}$
$<$ & 1.7 \\
MEKAL$^{\S{1}}$ & 0.38$^{+0.17}_{-0.16}$ \tablenotemark{g} & 20.0$^{<}_{-13.0}$ & N/A &
 0.3$^{+0.2}_{-0.2}$ & 0.2$^{+1.13}_{-0.13}$ & 2.8 \\
\hline
\end{tabular}
 
}
\tablenotetext{\S{1}\S{2}} {The spectral models are fitted
simultaneously with the models labeled by the same number.}
\tablenotetext{a} { 1 keV $\simeq$ 1.2 $\times 10^7 K$.}
\tablenotetext{b} { Calculated using the normalization constant of the
(V)MEKAL/(V)PSHOCK models. A = (10$^{-14}$/4$\pi$D$^2$)$\times$E.M. where
E.M. (Emission Measure) = $\int n_e\ n_H\ dV$ (integration is over the emitting
volume V).}    
\tablenotetext{c} {Unabsorbed Hard X-ray flux between 0.3 and 10 keV in
units of $\times 10^{-13}$ erg cm$^{-2}$ s$^{-1}$. The unconstrained parameters
are kept at their best fit values when calculating model fluxes.}
\tablenotetext{d} {This value of N$_H$ is derived from the fits with two different absorption parameter
used for the two X-ray spectral components.}
\tablenotetext{e} {The range is derived using the values for the two
parameters $\tau_{up}$ and $\tau_{low}$. $\tau_{low}$ is fixed at the value of
0.0 for convenience, since it is the lower limit for the integration over the
ionized volume and $\tau_{up}$ shows the maximum limit on the ionization
parameter at 2$\sigma$ confidence level.}
\tablenotetext{f} {The values  for $\chi^2_{\nu}$ are  listed with the
parameter results for Component 2
since the two emission models are fitted simultaneously.}
\tablenotetext{g} {This value of N$_H$ is derived from the fits with a single absorption parameter
used for both X-ray spectral components.}
\end{table}
 
\end{rotate}

\newpage
                                                                             
\figcaption{The X-ray nebula around the classical nova Persei (1901)
between 0.3 and 10 keV obtained with the \cha ACIS-S (S3).
The image shows emission 2${\sigma}$ above the background. The resolution
is 0$^{\prime\prime}$.5 per pixel. It is smoothed using a variable Gaussian 
filter with ${\sigma}$=0$^{\prime\prime}$.5-1$^{\prime\prime}$. 
North is up and West is to the right.
Central source is extracted from the image using a modeled PSF (see section 4.1) 
created with the aid of ChaRT, and MARX.}  

\figcaption{The X-ray data of the shell fitted with the
VMEKAL+PSHOCK emission models
including two different neutral hydrogen absorption. The smooth curve is
the \cha  ACIS-S response to the composite
model spectrum and the actual ACIS-S data are indicated with crosses.
The dashed line shows the contribution of the VMEKAL model whereas the
dot-dashed line
indicates the contribution from the PSHOCK model of emission.
The lower panel shows the residuals between the data and the
model in standard deviations. In general, solar elemental
abundances are assumed except for neon and nitrogen.}

\figcaption{(a) The source count rates extracted from an aperture of
12.$^{\prime\prime}$5 versus the source count rates detected within an annular
region from 12$^{\prime\prime}$.5 to 67$^{\prime\prime}$.5 in radius.
It shows a scaling relation between the source count rates (largely piled-up) and
the count rates derived from the PSF wings in the 2$^{\prime}$ vicinity of the pointed 
source. Both count rates are background subtracted. 
(b) The example calibration point
source spectra used in Figure 3a. The central source count rate (pileup rate)
increases from 0.19 to 0.21 c s$^{-1}$ from row 1 column 1 to row 2 col 2. The upper curves are the spectra from the
{\it source}-region and the lower curves are spectra from the {\it shell}-region as denoted in the text.} 

\figcaption{The top crosses (curve) indicate the data
of the central source spectrum derived from 
an aperture of 12$^{\prime\prime}$.5 (includes pileup photons). The middle crosses (curve)
show the simulated central source 
spectrum scaled to match the expected count rate from the 
location covered by the X-ray shell which is 0.037 c s$^{-1}$. The middle
curve is the expected spectrum of the PSF wings for GK Per. The lower crosses (curve) is the
un-piled central source spectrum (of GK Per) 
derived from the out-of-time events of the ACIS transfer streak.}
 
\figcaption{(a) The X-ray data of the shell cleaned from the effects
of the PSF wings of the central source  and fitted with a single model
of emission (VPSHOCK) together with a neutral hydrogen absorption model.
The figure stresses the significant deviation above 1.6 keV.
(b) The X-ray data of the shell cleaned from the effects of the
PSF wings of the central source  and fitted with the
VPSHOCK+NEI emission models
including two different neutral hydrogen absorption models. The smooth curve is
the \cha ACIS-S response to the composite
model spectrum and the actual ACIS-S data are indicated with crosses.
The dashed line shows the contribution of the VPSHOCK model whereas the
dot-dashed line
indicates the contribution from the NEI model of emission.
The lower panel shows the residuals between the data and the
model in standard deviations. (c) The same plot as in (b), but has only a single
neutral hydrogen absorption parameter for both of the components. 
This spectrum is regrouped to have a Signal-to-Noise ratio of 13 per energy bin.}

\figcaption{(a) The figure shows several extraction regions
used to derive the spectra to study the spectral variation within the remnant. 
The large circle is divided into 4 quadrants of SW, NW, NE, SE. The other three
regions are constructed to study the brightest X-ray region. The extraction 
region is a PIE (annulus-sector as noted in the text) with (1) denoting 
the inner ridge, (2) the Peak X-ray zone, and (3)  the outer ridge.
(b) The X-ray data fitted with the
VMEKAL+PSHOCK emission models
including neutral hydrogen absorption. The figure shows four spectra
obtained by dividing the shell into four quadrants centered on the
NE, SE, NW and SW with an inner radius of 12$^{\prime\prime}$.5 and an outer
radius  of 67$^{\prime\prime}$.5.}  

\figcaption{ (a) The X-ray data fitted with a single model of emission
(VPSHOCK) together with a single neutral hydrogen absorption model.
This figure stresses the significant deviation of the residuals above 1.6 keV
and the existence of the second component.
(b) The X-ray data fitted with the
VPSHOCK+SRCUT emission models
including two different neutral hydrogen absorption. The smooth curve is
the \cha ACIS-S response to the composite
model spectrum and the actual ACIS-S data are indicated with crosses.
The dashed line shows the contribution of the VPSHOCK model whereas the
dot-dashed line
indicates the contribution from the SRCUT model of emission.
The lower panel shows the residuals between the data and the
model in standard deviations.  (c) The same plot as in (b), but has only a single
neutral hydrogen absorption parameter for both of the components. }

\figcaption { The image of the X-ray nebula extracted from 
0.8 to 1.0 keV around the He-like Ne IX emission line. A suitable continuum
between energy channels 1.12-1.4 (corresponding to the amount of
energy channels within the line at the continuum level) is subtracted from the image.
The resolution of the image is
1$^{\prime\prime}$ per pixel. North is up and West is to the right.
The image is smoothed using a
Gaussian filter with  $\sigma$=2$^{\prime\prime}$. The cross point shows the location
of the central source.} 

\figcaption{ (a) The left-hand panel is the image of the X-ray nebula in the 0.25-0.50 keV
energy band, (b) The middle panel is the image of the X-ray nebula in  the 0.51-1.50 keV
energy band.  The two images in (a) $\&$ (b) show two sigma emission above the background at 
1$^{\prime\prime}$/pixel resolution, smoothed using a Gaussian filter with a variable 
$\sigma$=1$^{\prime\prime}$-2$^{\prime\prime}$. 
(c) The right-hand panel is the image of the X-ray nebula in  the 1.51-8.0 keV energy band 
at a lower resolution of 3$^{\prime\prime}$/pixel (compared with  (a) $\&$ (b)) 
smoothed using a Gaussian filter with a 
$\sigma$=6$^{\prime\prime}$. The image in (c) shows three sigma emission above the background.
The overlay are contours of the 0.51-1.50 keV image produced at the same spatial resolution and
smoothed using the same Gaussian.}

\figcaption{ (a) The figure shows the intensity contours of the HST image (1997) obtained
with the [NII] filter overlaid on the X-ray image of shell of GK Per. Both the HST contours and the X-ray
image show emission 2$\sigma$ above the background. 
The resolution of the combined image is 
2$^{\prime\prime}$ per pixel. Both images are also smoothed by a
Gaussian filter with $\sigma$=2$^{\prime\prime}$. 
The [NII] intensity contours are based on histogram equalization starting with
0.6$\sigma$ to 68$\sigma$ increments between contour levels. (b) 
The figure shows the radio intensity contours at 1.425 GHz
overlaid
on the X-ray image of the shell of nova Persei (1901). 
The  X-ray image and the radio contours  show
emission 2$\sigma$ and 1.5$\sigma$ above the background, respectively.
The resolution of the combined image is 2$^{\prime\prime}$ per pixel.
Both images are also smoothed by a
Gaussian filter with  $\sigma$=1$^{\prime\prime}$. The radio intensity contours are
constructed with linear 1.4$\sigma$ increments between contour levels.} 

\begin{center}

\begin{figure}
\plotone{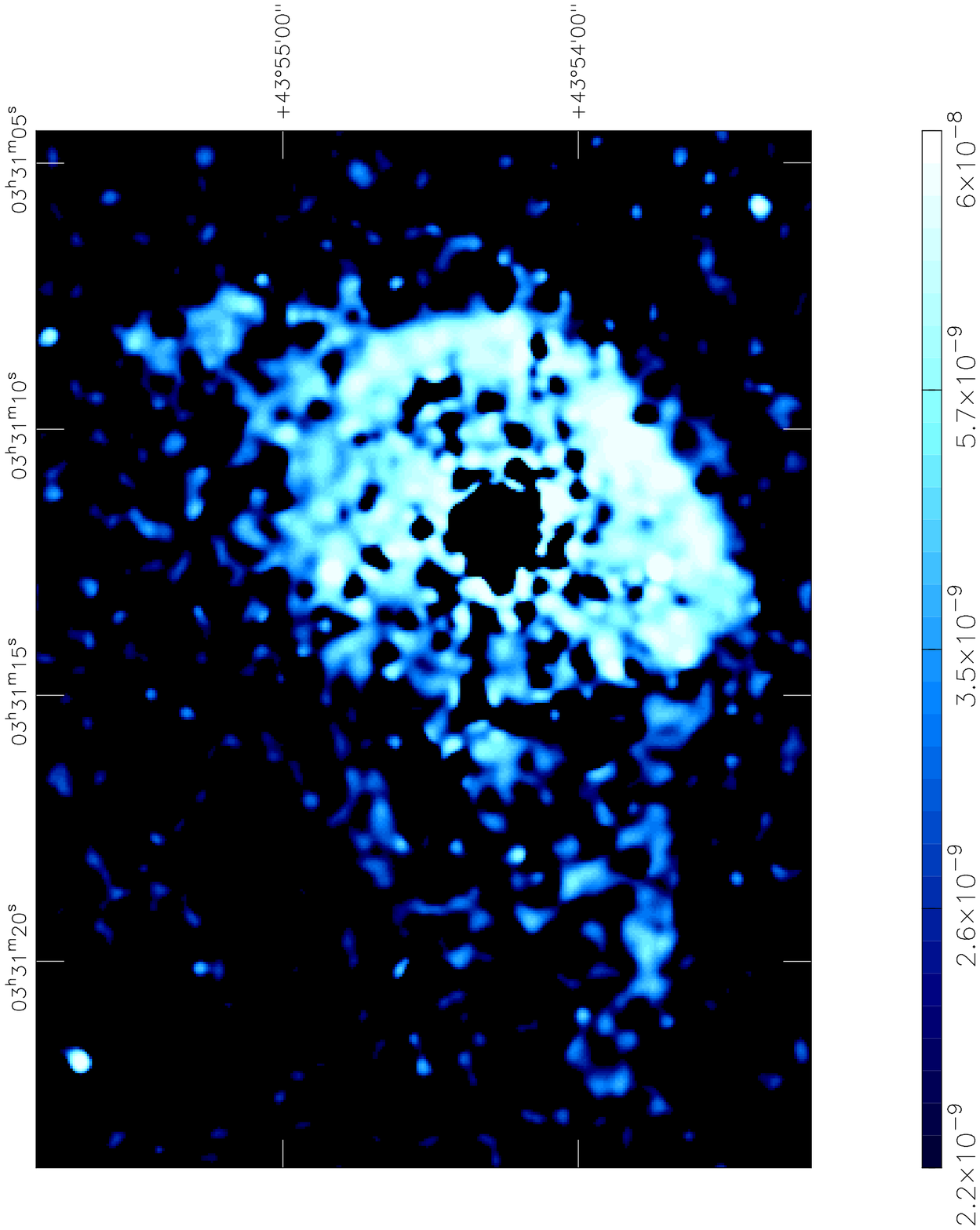}
\end{figure}                                                                                           
\begin{figure}
\plotone{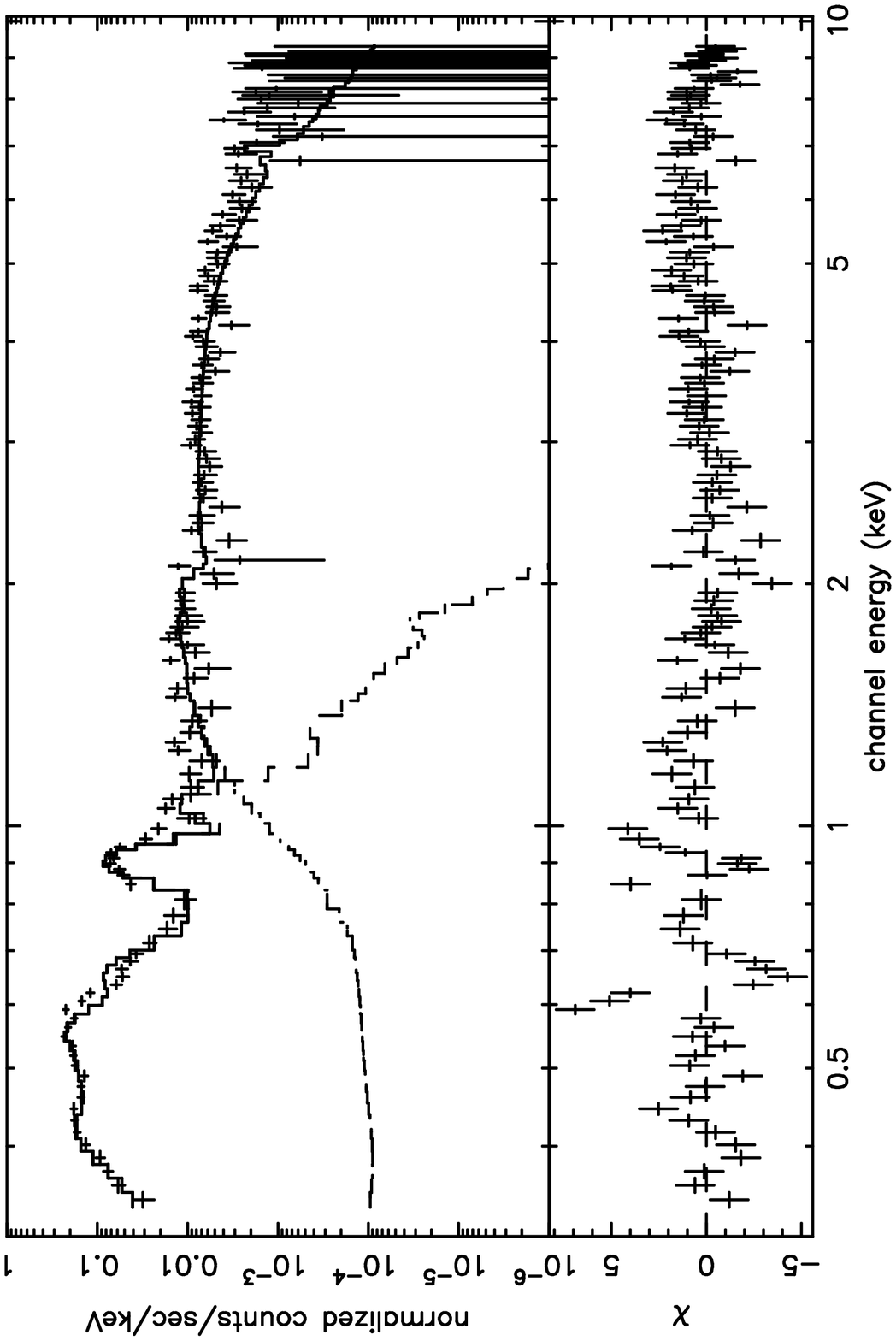}
\end{figure}        

\begin{figure}
\plotone{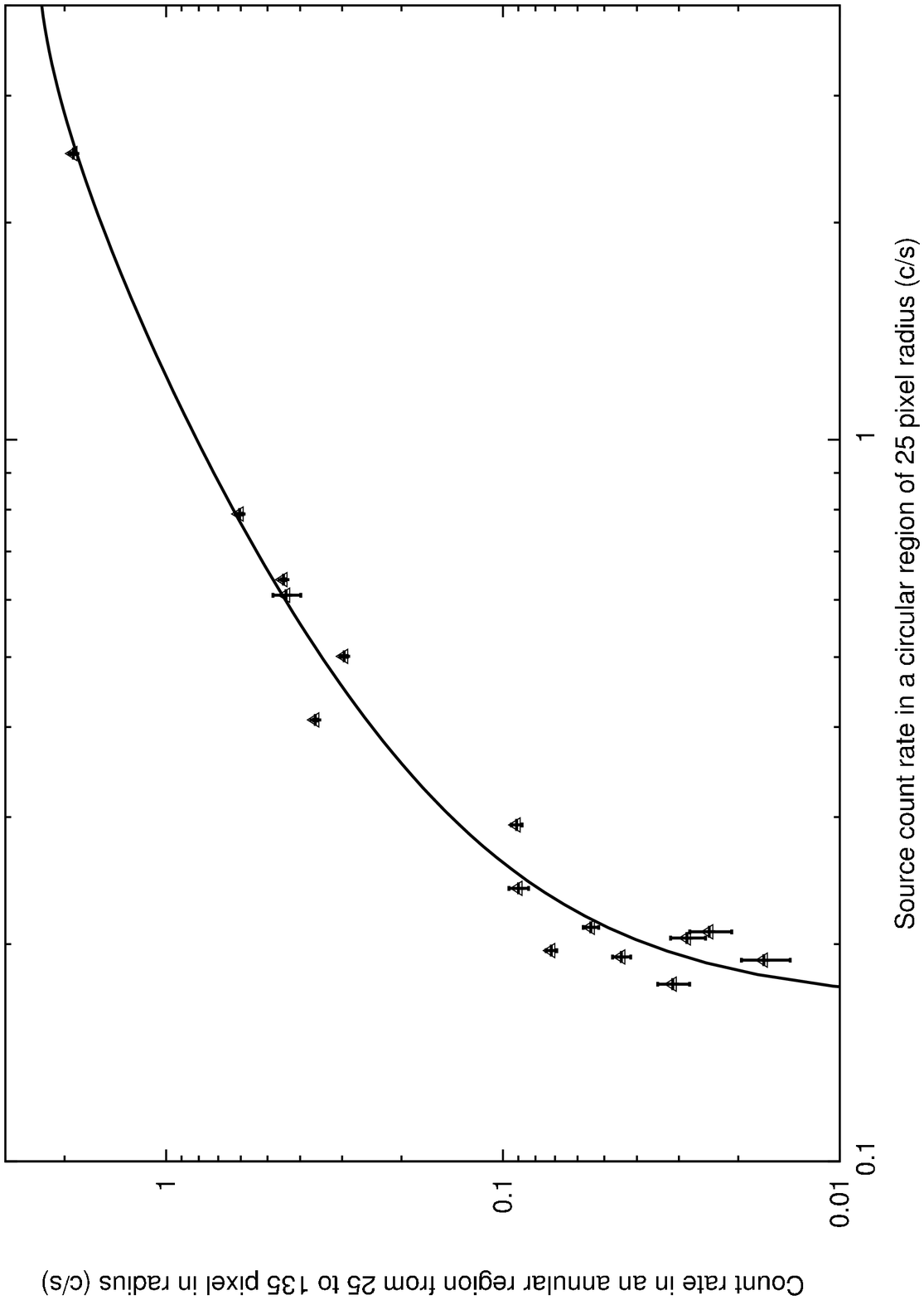}
\end{figure}

\begin{figure}
\includegraphics{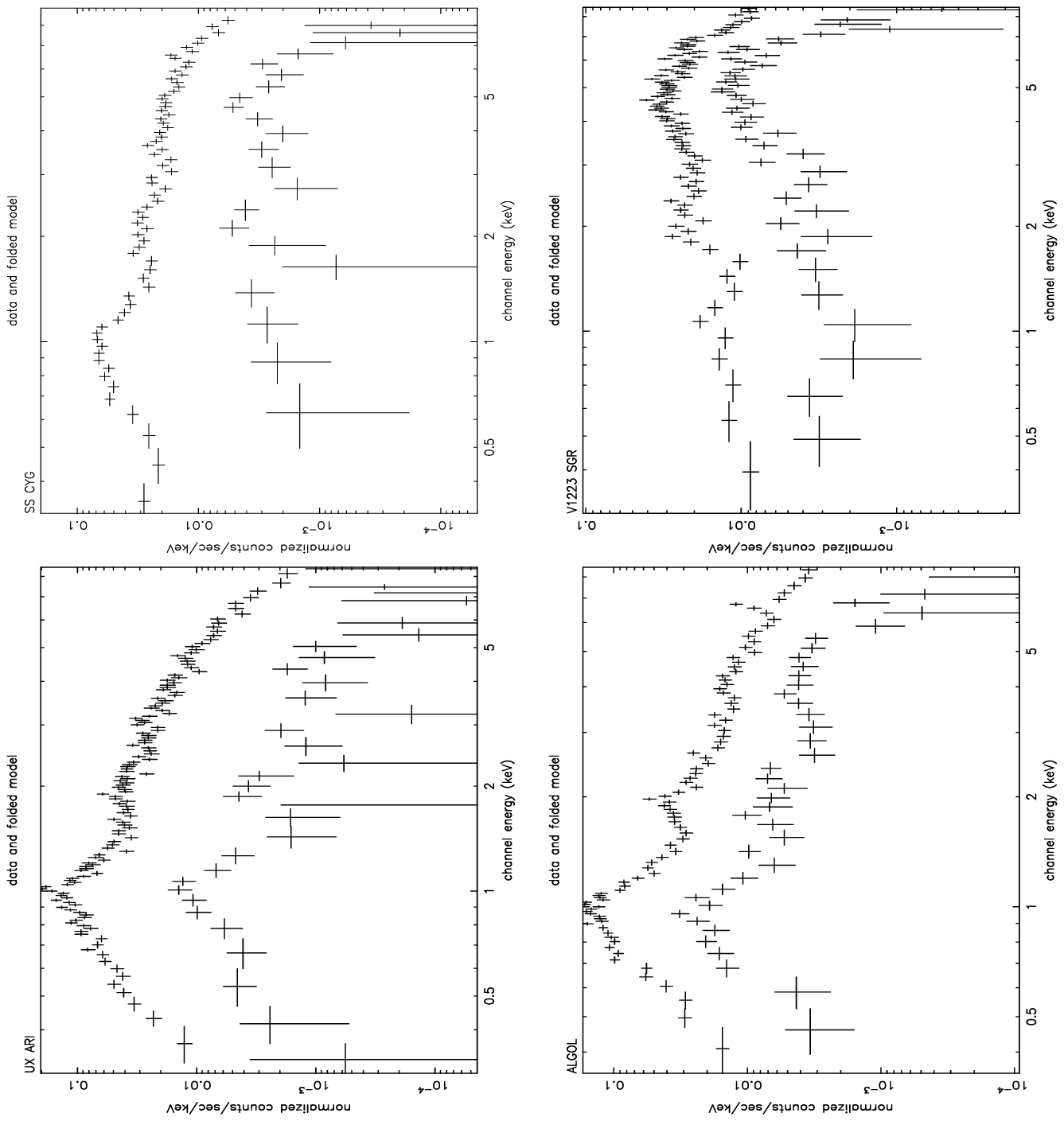}
\end{figure}

\begin{figure}
\plotone{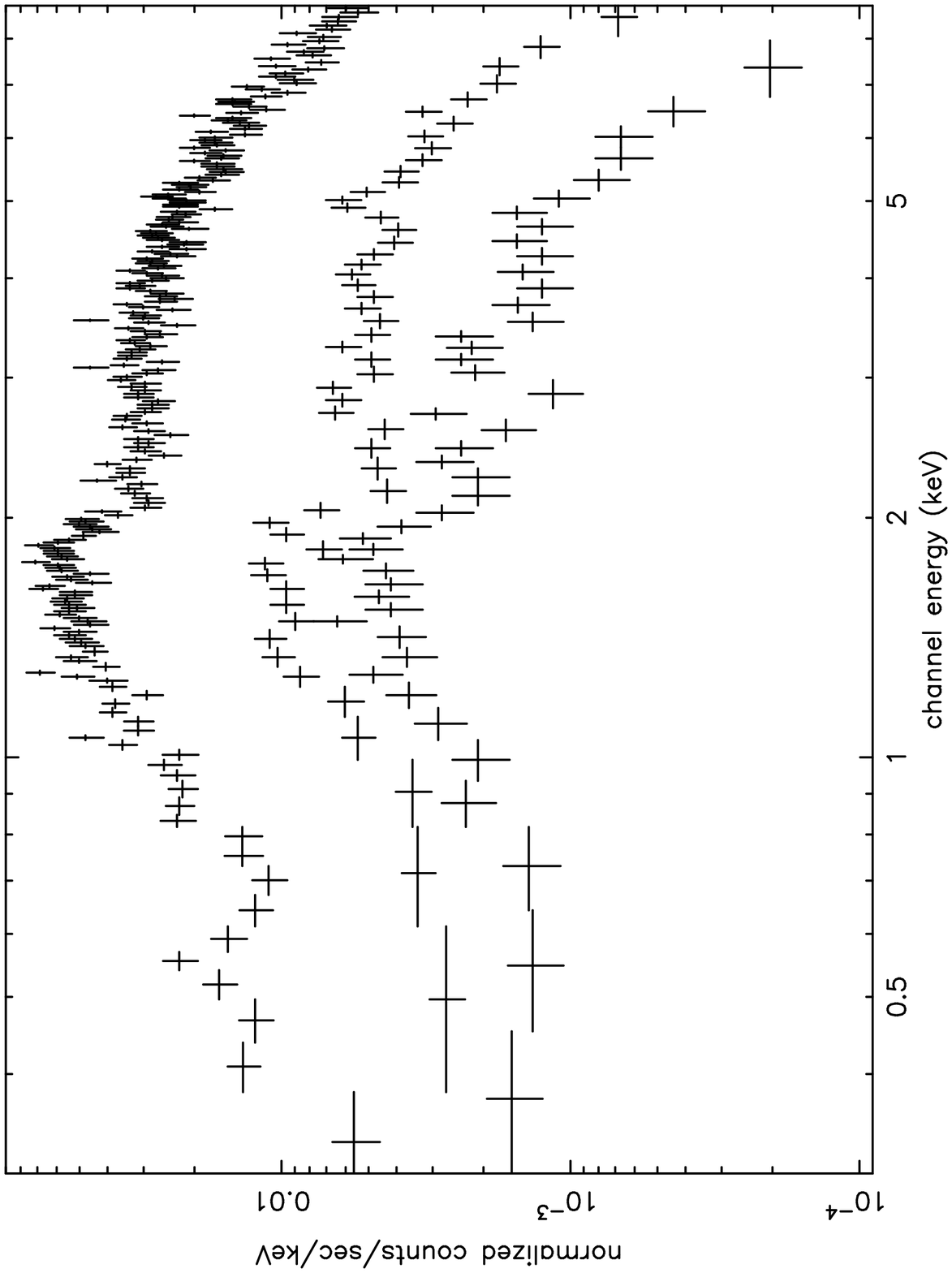}
\end{figure} 

\begin{figure}
\plotone{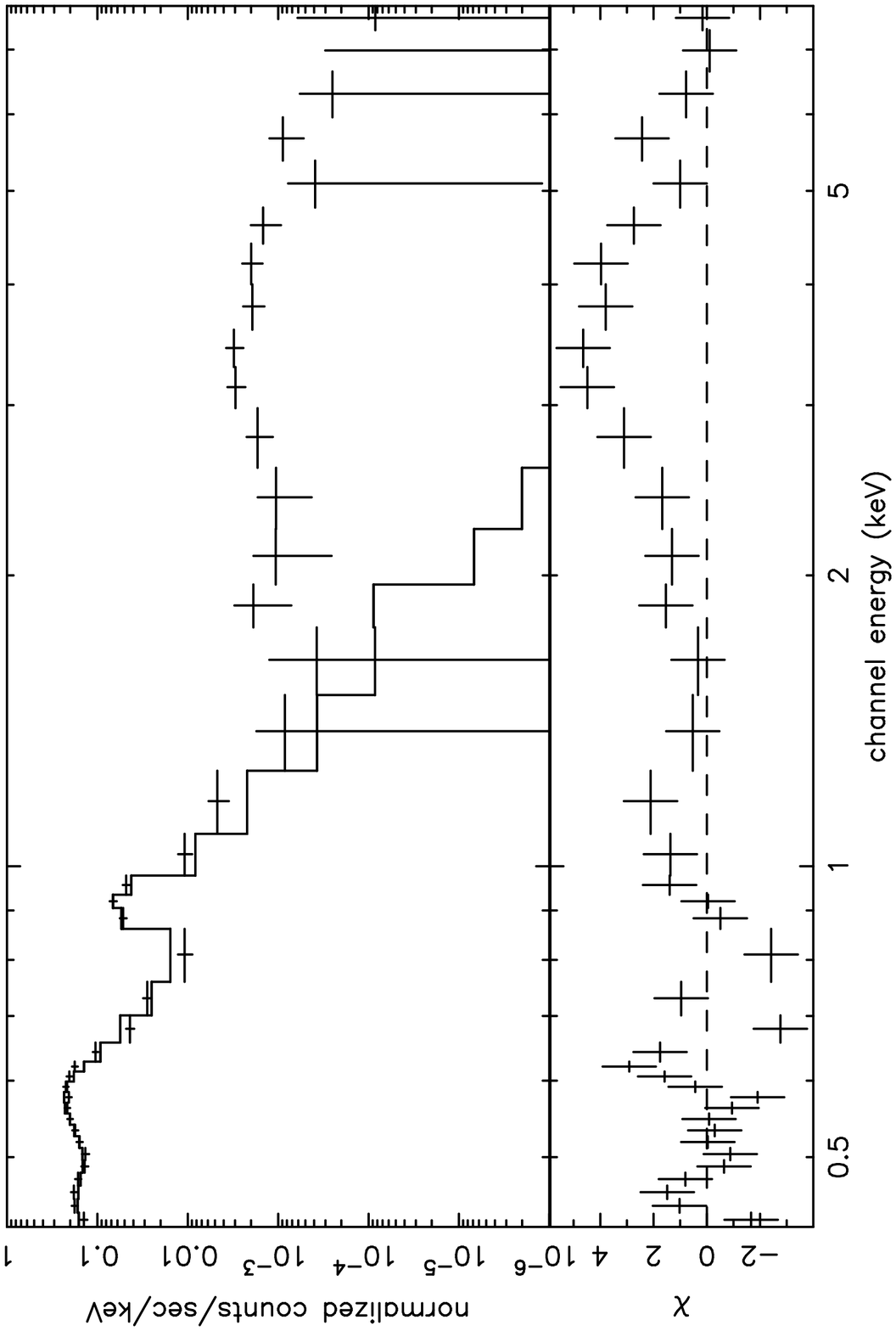}
\end{figure}                

\begin{figure}
\plotone{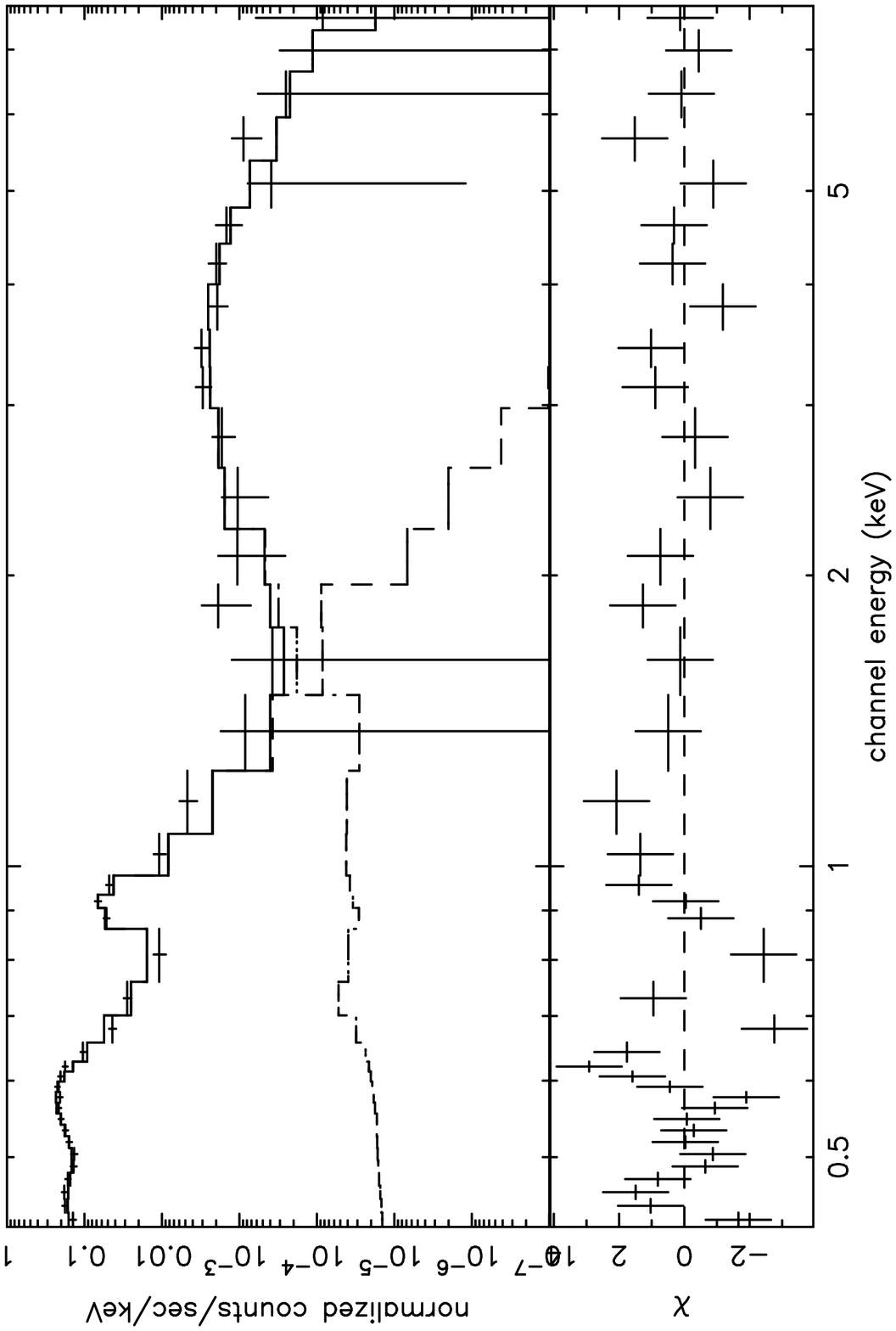}
\end{figure}

\begin{figure}
\plotone{f5c.ps}
\end{figure}

\begin{figure}
\plotone{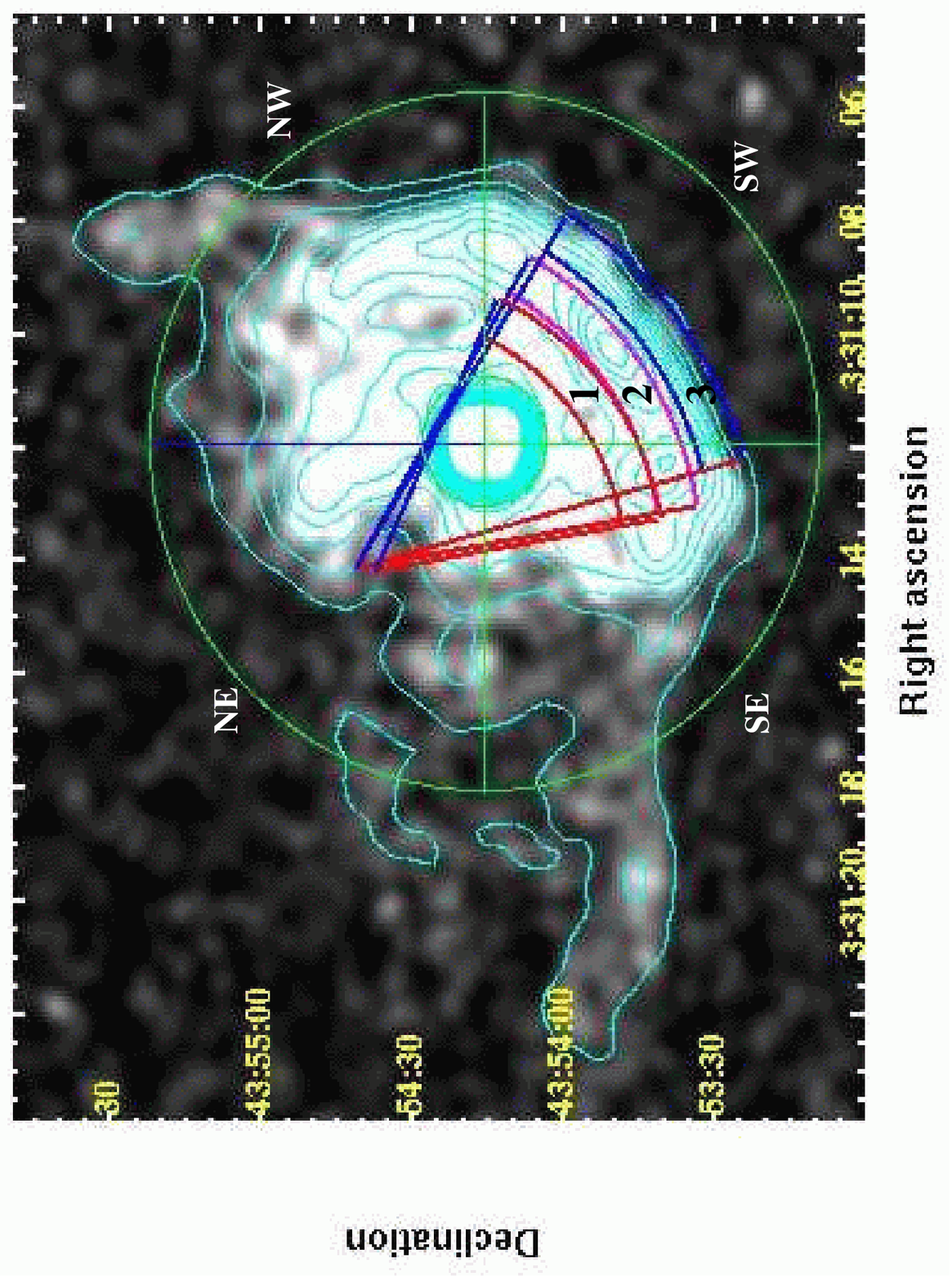}
\end{figure}

\begin{figure}
\plotone{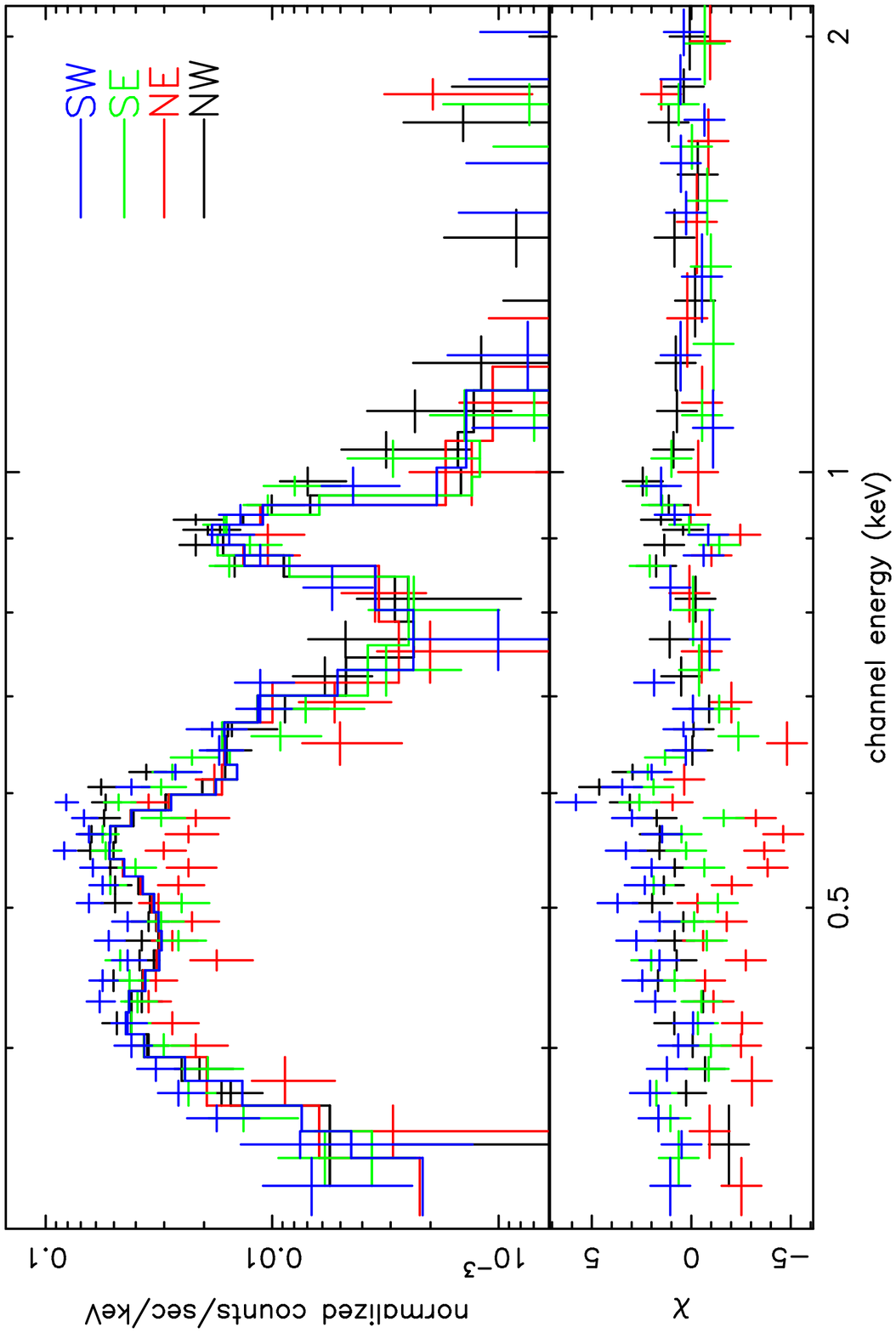}
\end{figure}             

\begin{figure}
\plotone{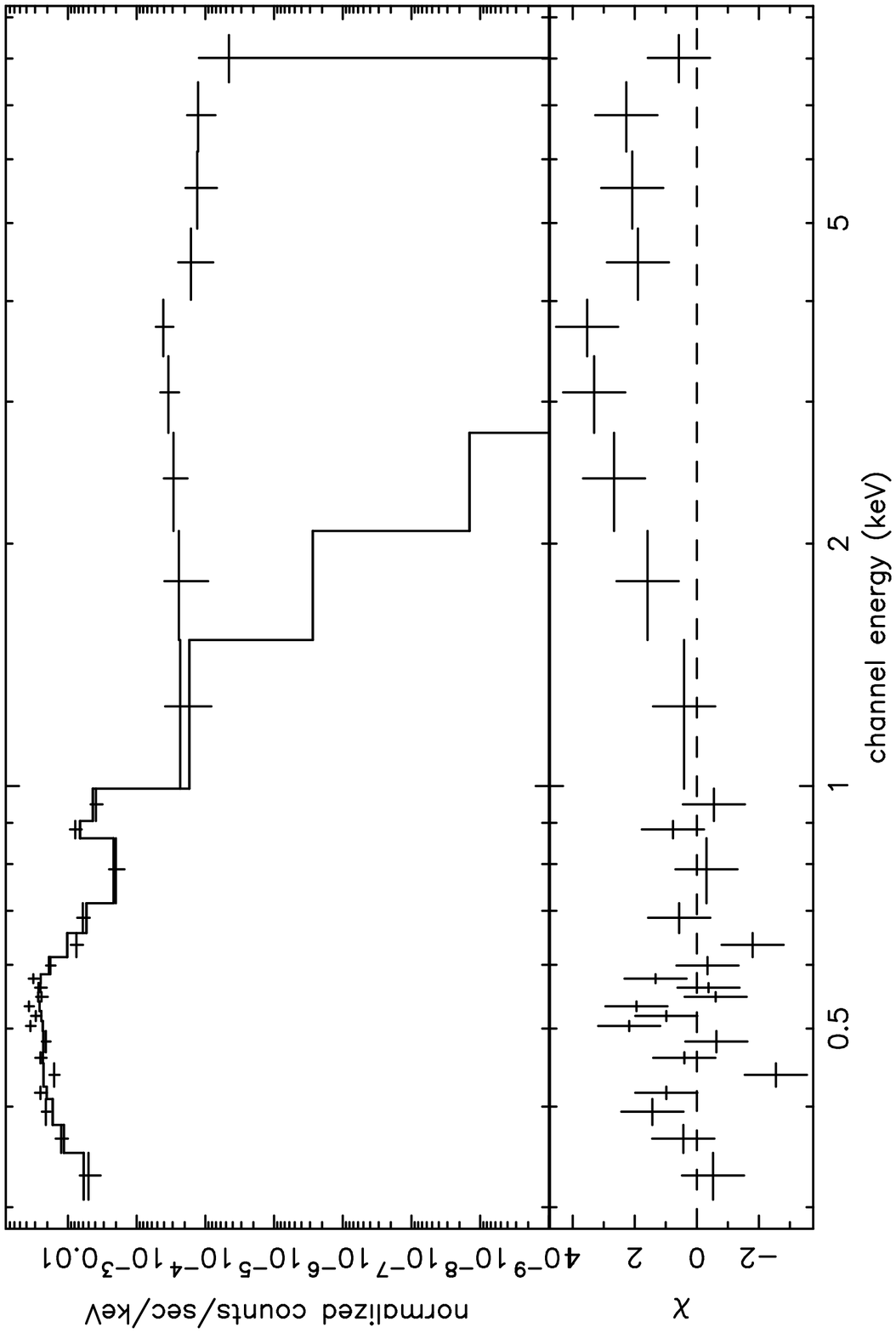}
\end{figure}     

\begin{figure}
\plotone{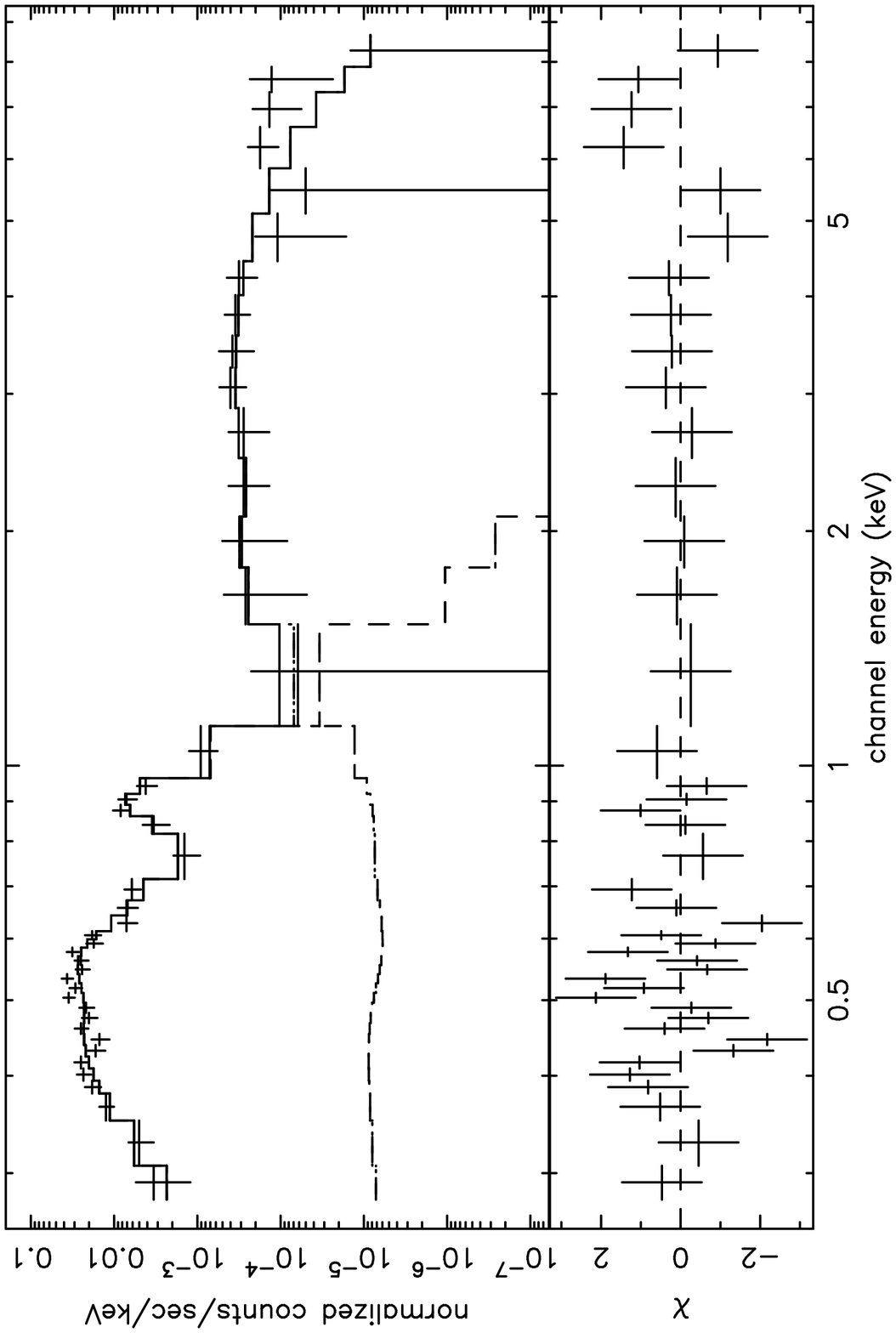}
\end{figure}

\begin{figure}
\plotone{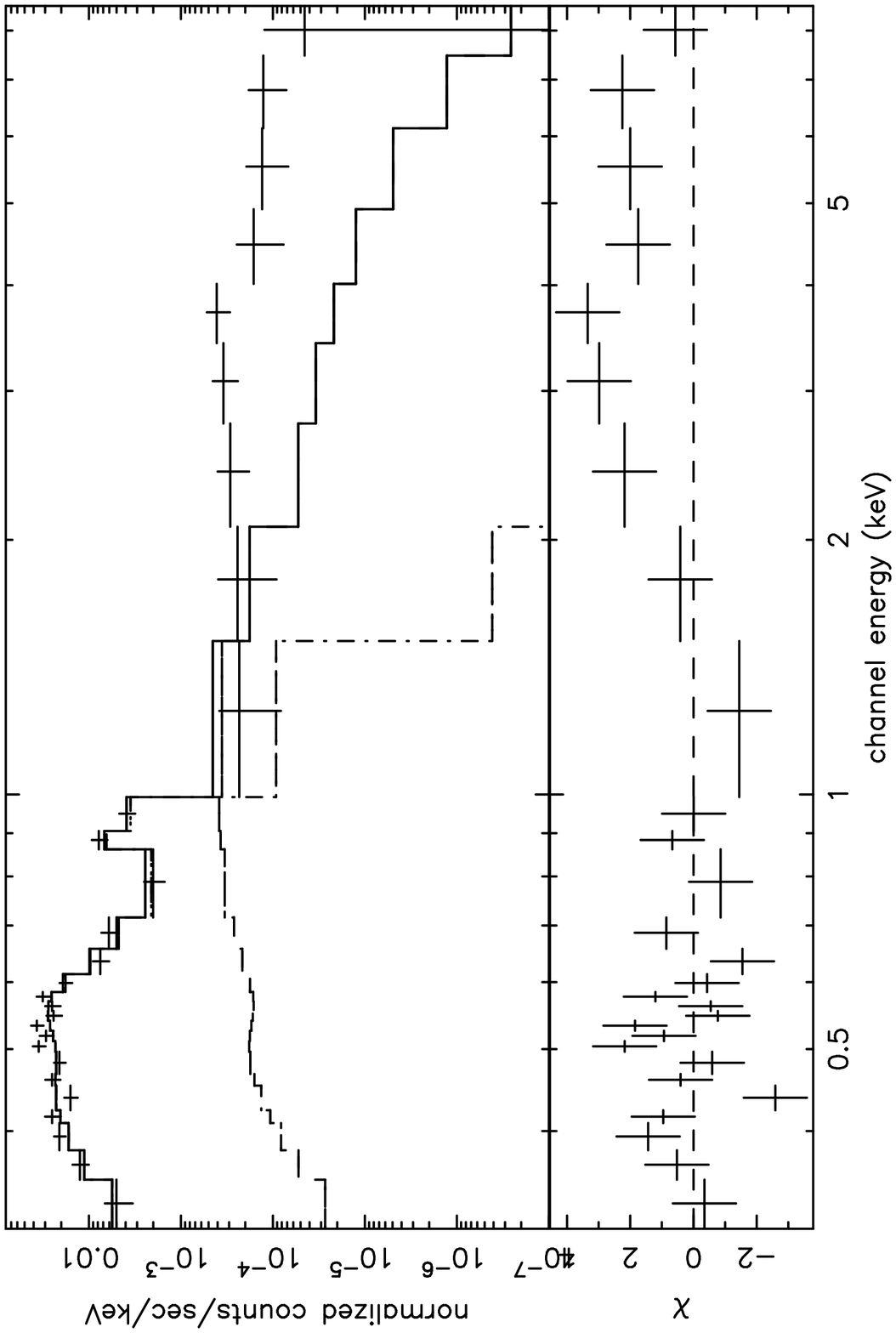}
\end{figure}

\begin{figure}
\plotone{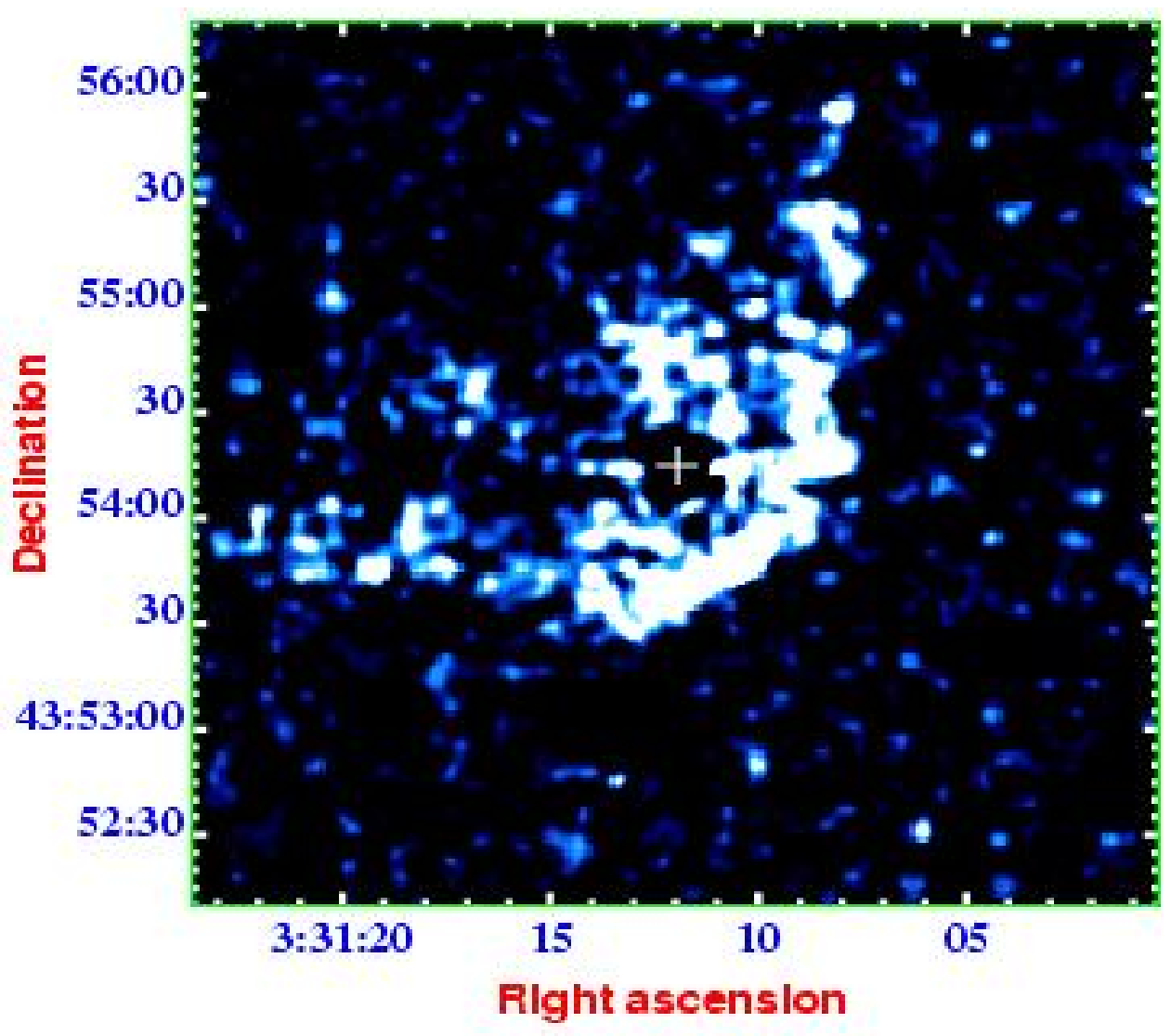}
\end{figure}  

\begin{figure}
\includegraphics{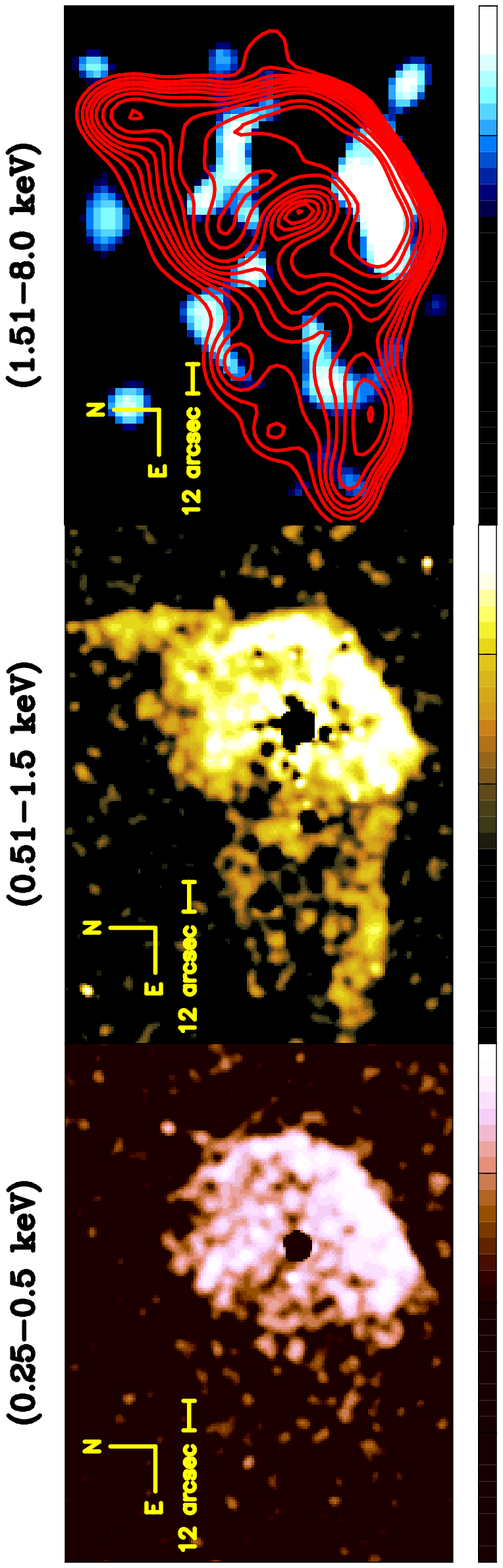}
\end{figure}         

\begin{figure}
\includegraphics{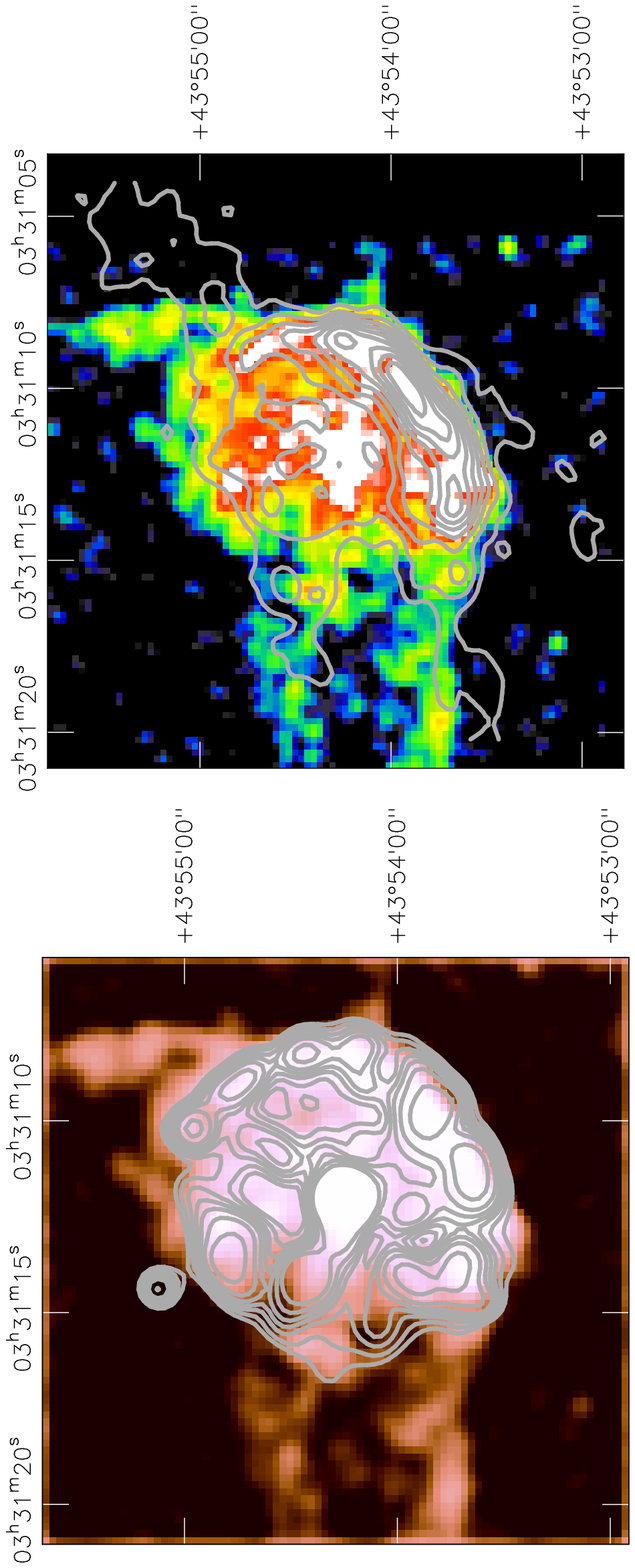}
\end{figure}

\end{center}
\end{document}